\documentclass[]{elsart}
\usepackage{amssymb}
\usepackage{graphicx,amsmath,amssymb}
\usepackage[below]{placeins}
\newcommand{\trD}[1]{\mbox{\boldmath $#1$}}
\newcommand{\vers}[1]{\hat{\trD{#1}}}

\newcommand{\spinoru}[2]{u_{#1}(\trD{#2})}

\newcommand{\spinorv}[2]{v_{#1}(\trD{#2})}

\newcommand{\Vpi}[3]{V_\pi^{#1}(\trD{#2},\trD{#3})}
\newcommand{\Vro}[3]{V_\rho^{#1}(\trD{#2},\trD{#3})}

\newcommand{\taut}[1]{\tau_{#1}\cdot\tau_\pi}

\begin{document}
\begin{frontmatter}
\title{
Microscopic derivation of pion-nucleon and pion-Delta scattering lengths
}
\author{P. Bicudo}\footnote{bicudo@ist.utl.pt},
\author{M. Faria},
\author{G. M. Marques}\footnote{gmarques@cfif.ist.utl.pt}
\author{and J. E. Ribeiro}\footnote{emilio@netcabo.pt}
\address{
Centro de F\'{\i}sica das Interac\c c\~oes Fundamentais (CFIF), \\
Departamento de F\'{\i}sica, Instituto Superior T\'ecnico, \\
Av. Rovisco Pais, 1049-001 Lisboa, Portugal
}

\maketitle
\begin{abstract}
A general expression for the $\pi -N$ and $\pi-\Delta $ scattering
lengths is derived in the framework of a microscopic calculation.
Annihilation, negative energy wave-functions and spontaneous
chiral symmetry are included consistently. The point-like limit is
used to calculate the scattering lengths.
\end{abstract}

\begin{keyword}
\\ preprint FISIST/18-2002/CFIF
\PACS 11.30.Rd, 12.39.Jh, 13.75.Gx
\end{keyword}

\end{frontmatter}

\section{Introduction}
\label{introduction}

\par
In the last decade a global picture for low energy hadronic
physics has slowly emerged. Although the theory of hadronic
reactions, which one expects to be a consequence of QCD, remains
as challenging as ever, we possess a natural symmetry (chiral
symmetry) which acts, so to speak, as a filter for  the still
largely unknown low energy details of strong interactions. Indeed
it is remarkable that although intermediate theoretical concepts
like gluon propagators, quark effective masses and so on, might
vary (in fact they are not gauge invariant and hence they are not
physical observables), chiral symmetry contrives for the final
physical results, e.g. hadronic masses and scattering lengths, to
be largely insensitive to the above mentioned theoretical
uncertainties. The pion mass furnishes the ultimate example. For
massless quarks, the pion mass is bound to be zero, regardless of
the form of the effective quark interaction {\em provided} it
supports the mechanism of spontaneous breakdown of chiral symmetry
(S$\chi$SB).

\par
$\pi -\pi$ elastic scattering
\cite{Weinberg,pi-pi,Bicudo1,Ribeiro2}
provides
another example of this insensitivity to  the form of the quark
microscopic interaction. The incorporation of chiral symmetry in
quark models has, in the history of hadronic physics, a long
standing, starting with the attempts at restoring chiral symmetry
to the MIT bag model
\cite{MIT,CT,Vento,CBMorig}.
Those
models can be considered as realizations  of a more general class
of microscopic models, known as extended Nambu--Jonas-Lasinio
\cite{Nambu}
models(eNJL)
\cite{LeYaou-pot,LeYaou-mes,Adler2,BicRib1,BicRib2,Felipe}
, which in turn could be formally deduced  from QCD
through cumulant expansion
\cite{Dosch}
if we were to know all the
gluon correlators. However, for practical applications, the set of
all gluon correlators gets reduced to the Gaussian approximation
(two gluon correlators). It simply turns out that chiral symmetry
alleviates us from the burden of knowing in detail even this
correlator. In fact, S$\chi$SB forces the quark-quark,
quark-antiquark, antiquark-antiquark potentials together with the
annihilation and creation interactions to originate from the
\emph{same} single chiral invariant Bethe-Salpeter kernel. Then it
turns out, in what concerns $\pi-$hadron scattering lengths, that
even this dependence is not \emph{explicit} so that such
scattering lengths are purely given in terms of hadronic masses
and normalizations. Of course the implicit dependence on a given
kernel is ensured by the actual values of those masses and
normalizations for that kernel. But what is important to stress is
that the formulae for the scattering lengths will not explicitly
depend on such kernels and will therefore remain formally
invariant for arbitrary kernels. The initial studies within eNJL
models were aimed at studying the interplay between confinement
and dynamical chiral symmetry breaking (D$\chi$SB), and
concentrated on critical couplings
\cite{LeYaou-pot}
for D$\chi$SB
and light-meson spectroscopy
\cite{LeYaou-mes}. Those NJL like
models have been extended to study the pion beyond BCS level and
meson resonant decays in the context of a generalized
\cite{BicRib2,Bicudo,Bicudo4,Bicudo8,Bicudo9,Bicudo10}
resonating group method (RGM)
\cite{Ribeiro1}.

\par
In this paper we will derive the scattering lengths for the
$\pi-B$ system, with $B$ standing for either $N$ or $\Delta$. This
notation will be used throughout this paper. The effects of
quark-antiquark annihilation, of the relativistic negative energy
component of the pion, and of the hadron exchange interaction will
be consistently computed in the framework of the S$\chi$SB.
This will also improve the state of the art
\cite{Lemaire,Ceuleneer}
of the RGM.

\par
This paper is organized as follows: in Section \ref{Hamiltonian}
we present the Hamiltonian and some useful definitions; in Section
\ref{RGM} we determine the RGM diagrams which contribute for the
contact effective pion-hadron potential; Section \ref{bound state}
is devoted to the the study of the Salpeter amplitudes of mesons
and baryons; in Section \ref{assembling} we compute the various
contributions to $\mathcal{O}^{\mbox{\tiny RGM}}_{\pi-B}$ and
explain how to assemble them together; the computation of the one
hadron exchange effective potential can be avoided in the
point-like limit; in Section \ref{just contact} the full effective
potential is computed just with the contact term; we conclude in
Section \ref{conclusion}. The technical details involving flavor
and spin traces are left for Appendix \ref{traces}.

\section{The Hamiltonian}
\label{Hamiltonian}

The eNJL class of models correspond to the Gaussian approximation
of QCD in the cumulant expansion in terms of gluon correlators,
\begin{eqnarray}
H  &=& \int d^3x \, q^\dagger(x)(-i \trD{\alpha}\cdot\trD{\nabla}+m\beta) q(x)
\nonumber\\
& & - \frac{i}{2} \int d^3x \int d^3y \, \bar{q}(x) \gamma^\mu  T^a q(x)
\bar{q}(y) \gamma^\nu T^b q(y) \langle\langle g A^a_\mu(x) g A^b_\nu(y)
\rangle\rangle,
\label{heff}
\end{eqnarray}
The Gaussian correlator $\langle \langle g A^a_\mu(x) g
A^b_\nu(y)\rangle \rangle$ turns out to be identical to the
va\-cuum expectation of $K(x,y)=\langle g A^a_\mu(x) g
A^b_\nu(y)\rangle$. Translation invariance, $K(x,y)=K(x-y)$ is
assumed throughout this paper. In Eq. (\ref{heff}) the $T^a$
stands for the Gell-Mann generator $\lambda^a /2$ and the quark
field $q(\trD{x})$ is given by

\begin{equation}
q(\trD{x})=\sum_s\int {d^3 k \over (2 \pi)^3} \left[
u_s(\trD{k})b_{{\bf k}\, s} + v_s(\trD{k})d^\dagger_{-{\bf k}\, s}
\right] e^{i {\bf k} \cdot {\bf x}} \ ,
\end{equation}
with
\begin{equation} \label{eq:spinors}
\begin{split}
u_s(\trD{k})&=\left[
\sqrt{\frac{1+S(k)}{2}} + \sqrt{\frac{1-S(k)}{2}}\ \vers{k} \cdot \trD{\alpha}
\right]u_s(0) \ , \\
v_s(\trD{k})&=\left[ \sqrt{\frac{1+S(k)}{2}} -
\sqrt{\frac{1-S(k)}{2}}\ \vers{k} \cdot \trD{\alpha} \right]v_s(0)
\ .
\end{split}
\end{equation}
For the quark spinors $\spinoru{\uparrow}{0}=(1,0,0,0)$,
$\spinoru{\downarrow}{0}=(0,1,0,0)$,
$\spinorv{\uparrow}{0}=(0,0,0,1)$ and
$\spinorv{\downarrow}{0}=(0,0,-1,0)$. In Eq. (\ref{eq:spinors})
$S(k)$ is a shorthand notation for $\sin\phi(k)$ the solution of
the S$\chi$SB mass gap equation. For details on the solution of
the mass gap equation see Ref.
\cite{LeYaou-pot,Adler2,BicRib1,Bicudo6}. Following this reference
we arrive at the set of possible quark-quark, quark-antiquark and
antiquark-antiquark vertices, which are defined in Table
\ref{vertices}, with the convention of representing the quark with
a single line and the antiquark with a double line. For
completeness we also represent in the last entry the single-quark
(antiquark) energy $E(p)$, as a function of the modulus of the
momentum $\trD{p}$.

\begin{table}
\caption{ The microscopic vertices for the coupling of the
potential rung with the quark and antiquark lines.}
\begin{tabular}{cccc}
\hline \hline Quark vertex & $\Gamma^{qq}$ & $u^\dagger_s(k) \beta
\gamma^\mu u_{s'}(k') $ &
\begin{picture}(20,5)(0,0)
\put(0,2){\line(1,0){20}} \multiput(10,.5)(0,-5){2}{\line(0,1){2}}
\put(10,2){\circle*{2}}
\end{picture}
\\
Antiquark vertex & $\Gamma^{\bar q \bar q}$ & $-v^\dagger_s(k)
\beta \gamma^\mu v_{s'}(k') $ &
\begin{picture}(20,5)(0,0)
\put(0,1){\line(1,0){20}} \put(0,3){\line(1,0){20}}
\multiput(10,.5)(0,-5){2}{\line(0,1){2}} \put(10,2){\circle*{3}}
\end{picture}
\\
Creation vertex & $\Gamma^{q \bar q}$ & $u^\dagger_s(k) \beta
\gamma^\mu v_{s'}(k') $ &
\begin{picture}(20,5)(0,0)
\put(10,3){\line(-2,1){10}} \put(10,1){\line(-2,-1){10}}
\put(10,3){\line(-2,-1){10}}
\multiput(10,.5)(0,-5){2}{\line(0,1){2}} \put(10,2){\circle*{3}}
\end{picture}
\\
Annihilation vertex & $\Gamma^{\bar q q}$ & $v^\dagger_s(k) \beta
\gamma^\mu u_{s'}(k') $ &
\begin{picture}(20,5)(0,0)
\put(10,3){\line(2,1){10}} \put(10,1){\line(2,-1){10}}
\put(10,3){\line(2,-1){10}}
\multiput(10,.5)(0,-5){2}{\line(0,1){2}} \put(10,2){\circle*{3}}
\end{picture}
\\
Kinetic energy & $T_p$  &  &
\begin{picture}(20,5)(0,0)
\put(0,2){\line(1,0){20}} \put(10,2){\circle*{4}}
\end{picture}
\\
\hline
\end{tabular}
\label{vertices}
\end{table}

\begin{figure}
\centering
\includegraphics[width=1.\textwidth]{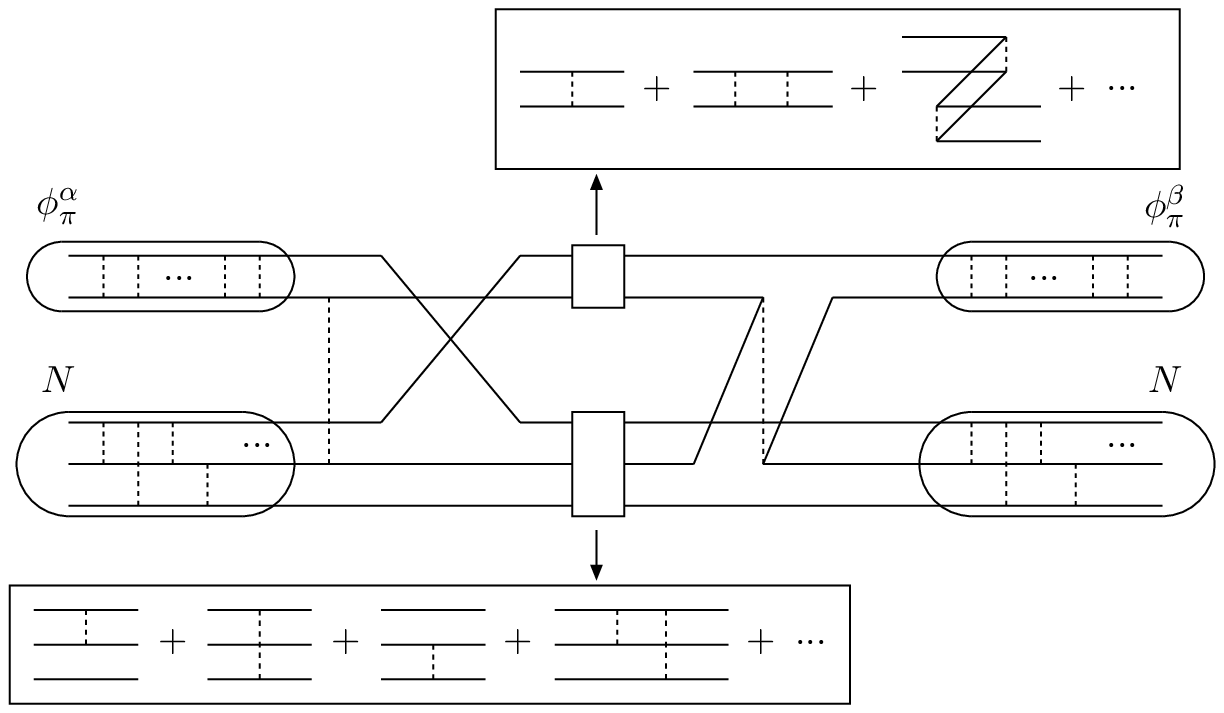}
\caption{ Example of a diagram with intermediate hadrons}
\label{intermhadron}
\end{figure}

The set of all diagrams which can be built from the basic vertices
of Table \ref{vertices} fall into two disjoint classes:
\begin{itemize}
\item Diagrams in which no intermediate hadron propagator can be
cut. As an example of diagrams belonging to this class see Fig.
\ref{RGM diagrams}. We call the diagrams belonging to this class
contact diagrams. \item The complementary class of diagrams having
an intermediate hadron. See Fig. \ref{intermhadron} for an
example.
\end{itemize}

In the remainder of his paper we will derive the $\pi-$Baryon
($\pi-B$) scattering lengths, $a_{\pi-B}$.

\section{The Resonating Group Method evaluation of $a_{\pi-B}$}
\label{RGM}

\begin{figure}
%
\begin{picture}(200,120)(0,0)
\put(20,5){
\begin{picture}(120,70)(0,0)
\put(-20,40){$\langle E \rangle = E \times$}
\put(20,0){
\begin{picture}(30,100)(0,0)
\put(10,8){$\phi$} \put(20,-5){\line(0,1){30}}
\put(0,10){\line(4,3){20}} \put(0,10){\line(4,-3){20}}
\put(10,70){$\psi$} \put(20,45){\line(0,1){50}}
\put(0,70){\line(4,5){20}} \put(0,70){\line(4,-5){20}}
\end{picture}}
\put(117.5,0){
\begin{picture}(30,100)(0,0)
\put(3,8){$\phi$} \put(0,-5){\line(0,1){30}}
\put(20,10){\line(-4,3){20}} \put(20,10){\line(-4,-3){20}}
\put(3,70){$\psi$} \put(0,45){\line(0,1){50}}
\put(20,70){\line(-4,5){20}} \put(20,70){\line(-4,-5){20}}
\end{picture}}
\put(40,0){
\begin{picture}(30,100)(0,0)
\put(0,-1){\line(1,0){30}} \put(0,1){\line(1,0){30}}
\put(0,20){\line(1,0){30}} \put(0,50){\line(1,0){30}}
\put(0,70){\line(1,0){30}} \put(0,90){\line(1,0){30}}
\end{picture}}
\put(87.5,0){
\begin{picture}(50,100)(0,0)
\put(0,-1){\line(1,0){30}} \put(0,1){\line(1,0){30}}
\put(0,20){\line(1,0){30}} \put(0,50){\line(1,0){30}}
\put(0,70){\line(1,0){30}} \put(0,90){\line(1,0){30}}
\end{picture}}
\put(53,0){
\begin{picture}(20,100)(0,0)
\put(0,10){\oval(20,14)[l]} \put(-10,10){\vector(0,-1){2}}
\put(-5,8){$p$} \put(0,60){\oval(20,14)[l]}
\put(-10,60){\vector(0,-1){2}} \put(-5,56){$k$}
\put(0,80){\oval(20,14)[l]} \put(-10,80){\vector(0,-1){2}}
\put(-5,78){$p$}
\end{picture}}
\put(105,0){
\begin{picture}(20,100)(0,0)
\put(0,10){\oval(20,14)[r]} \put(10,10){\vector(0,1){2}}
\put(0,8){$p$} \put(0,60){\oval(20,14)[r]}
\put(10,60){\vector(0,1){2}} \put(0,56){$k$}
\put(0,80){\oval(20,14)[r]} \put(10,80){\vector(0,1){2}}
\put(0,78){$p$}
\end{picture}}
\put(70,0){
\begin{picture}(20,100)(0,0)
\put(0,20){\line(1,4){17.5}} \put(0,90){\line(1,-4){17.5}}
\put(0,-1){\line(1,0){17.5}} \put(0,1){\line(1,0){17.5}}
\put(0,50){\line(1,0){17.5}} \put(0,70){\line(1,0){17.5}}
\end{picture}}
\end{picture}}
\end{picture}
%
\begin{picture}(200,120)(0,0)
\put(10,5){
\begin{picture}(120,70)(0,0)
\put(-10,40){$\langle T_1 \rangle =$}
\put(55,0){
\begin{picture}(20,100)(0,0)
\put(0,97){$T_p$}
\end{picture}}
\put(20,0){
\begin{picture}(30,100)(0,0)
\put(10,8){$\phi$} \put(20,-5){\line(0,1){30}}
\put(0,10){\line(4,3){20}} \put(0,10){\line(4,-3){20}}
\put(10,70){$\psi$} \put(20,45){\line(0,1){50}}
\put(0,70){\line(4,5){20}} \put(0,70){\line(4,-5){20}}
\end{picture}}
\put(137.5,0){
\begin{picture}(30,100)(0,0)
\put(3,8){$\phi$} \put(0,-5){\line(0,1){30}}
\put(20,10){\line(-4,3){20}} \put(20,10){\line(-4,-3){20}}
\put(3,70){$\psi$} \put(0,45){\line(0,1){50}}
\put(20,70){\line(-4,5){20}} \put(20,70){\line(-4,-5){20}}
\end{picture}}
\put(40,0){
\begin{picture}(30,100)(0,0)
\put(0,-1){\line(1,0){30}} \put(0,1){\line(1,0){30}}
\put(0,20){\line(1,0){30}} \put(0,50){\line(1,0){30}}
\put(0,70){\line(1,0){30}} \put(0,90){\line(1,0){30}}
\end{picture}}
\put(107.5,0){
\begin{picture}(50,100)(0,0)
\put(0,-1){\line(1,0){30}} \put(0,1){\line(1,0){30}}
\put(0,20){\line(1,0){30}} \put(0,50){\line(1,0){30}}
\put(0,70){\line(1,0){30}} \put(0,90){\line(1,0){30}}
\end{picture}}
\put(70,0){
\begin{picture}(20,100)(0,0)
\put(0,20){\line(1,0){20}} \put(0,90){\line(1,0){20}}
\put(0,-1){\line(1,0){20}} \put(0,1){\line(1,0){20}}
\put(0,50){\line(1,0){20}} \put(0,70){\line(1,0){20}}
\put(0,90){\circle*{5}}
\end{picture}}
\put(90,0){
\begin{picture}(20,100)(0,0)
\put(0,20){\line(1,4){17.5}} \put(0,90){\line(1,-4){17.5}}
\put(0,-1){\line(1,0){17.5}} \put(0,1){\line(1,0){17.5}}
\put(0,50){\line(1,0){17.5}} \put(0,70){\line(1,0){17.5}}
\end{picture}}
\end{picture}}
\end{picture}
%
\\
%
\begin{picture}(200,120)(0,0)
\put(10,5){
\begin{picture}(120,70)(0,0)
\put(-10,40){$\langle T_2 \rangle=$}
\put(20,0){
\begin{picture}(30,100)(0,0)
\put(10,8){$\phi$} \put(20,-5){\line(0,1){30}}
\put(0,10){\line(4,3){20}} \put(0,10){\line(4,-3){20}}
\put(10,70){$\psi$} \put(20,45){\line(0,1){50}}
\put(0,70){\line(4,5){20}} \put(0,70){\line(4,-5){20}}
\end{picture}}
\put(137.5,0){
\begin{picture}(30,100)(0,0)
\put(3,8){$\phi$} \put(0,-5){\line(0,1){30}}
\put(20,10){\line(-4,3){20}} \put(20,10){\line(-4,-3){20}}
\put(3,70){$\psi$} \put(0,45){\line(0,1){50}}
\put(20,70){\line(-4,5){20}} \put(20,70){\line(-4,-5){20}}
\end{picture}}
\put(40,0){
\begin{picture}(30,100)(0,0)
\put(0,-1){\line(1,0){30}} \put(0,1){\line(1,0){30}}
\put(0,20){\line(1,0){30}} \put(0,50){\line(1,0){30}}
\put(0,70){\line(1,0){30}} \put(0,90){\line(1,0){30}}
\end{picture}}
\put(107.5,0){
\begin{picture}(50,100)(0,0)
\put(0,-1){\line(1,0){30}} \put(0,1){\line(1,0){30}}
\put(0,20){\line(1,0){30}} \put(0,50){\line(1,0){30}}
\put(0,70){\line(1,0){30}} \put(0,90){\line(1,0){30}}
\end{picture}}
\put(70,0){
\begin{picture}(20,100)(0,0)
\put(0,20){\line(1,0){20}} \put(0,90){\line(1,0){20}}
\put(0,-1){\line(1,0){20}} \put(0,1){\line(1,0){20}}
\put(0,50){\line(1,0){20}} \put(0,70){\line(1,0){20}}
\put(0,70){\circle*{5}}
\end{picture}}
\put(90,0){
\begin{picture}(20,100)(0,0)
\put(0,20){\line(1,4){17.5}} \put(0,90){\line(1,-4){17.5}}
\put(0,-1){\line(1,0){17.5}} \put(0,1){\line(1,0){17.5}}
\put(0,50){\line(1,0){17.5}} \put(0,70){\line(1,0){17.5}}
\end{picture}}
\end{picture}}
\end{picture}
%
\begin{picture}(200,120)(0,0)
\put(10,5){
\begin{picture}(120,70)(0,0)
\put(-10,40){$\langle T_5 \rangle=$}
\put(20,0){
\begin{picture}(30,100)(0,0)
\put(10,8){$\phi$} \put(20,-5){\line(0,1){30}}
\put(0,10){\line(4,3){20}} \put(0,10){\line(4,-3){20}}
\put(10,70){$\psi$} \put(20,45){\line(0,1){50}}
\put(0,70){\line(4,5){20}} \put(0,70){\line(4,-5){20}}
\end{picture}}
\put(137.5,0){
\begin{picture}(30,100)(0,0)
\put(3,8){$\phi$} \put(0,-5){\line(0,1){30}}
\put(20,10){\line(-4,3){20}} \put(20,10){\line(-4,-3){20}}
\put(3,70){$\psi$} \put(0,45){\line(0,1){50}}
\put(20,70){\line(-4,5){20}} \put(20,70){\line(-4,-5){20}}
\end{picture}}
\put(40,0){
\begin{picture}(30,100)(0,0)
\put(0,-1){\line(1,0){30}} \put(0,1){\line(1,0){30}}
\put(0,20){\line(1,0){30}} \put(0,50){\line(1,0){30}}
\put(0,70){\line(1,0){30}} \put(0,90){\line(1,0){30}}
\end{picture}}
\put(107.5,0){
\begin{picture}(50,100)(0,0)
\put(0,-1){\line(1,0){30}} \put(0,1){\line(1,0){30}}
\put(0,20){\line(1,0){30}} \put(0,50){\line(1,0){30}}
\put(0,70){\line(1,0){30}} \put(0,90){\line(1,0){30}}
\end{picture}}
\put(70,0){
\begin{picture}(20,100)(0,0)
\put(0,20){\line(1,0){20}} \put(0,90){\line(1,0){20}}
\put(0,-1){\line(1,0){20}} \put(0,1){\line(1,0){20}}
\put(0,50){\line(1,0){20}} \put(0,70){\line(1,0){20}}
\put(0,0){\circle*{5}}
\end{picture}}
\put(90,0){
\begin{picture}(20,100)(0,0)
\put(0,20){\line(1,4){17.5}} \put(0,90){\line(1,-4){17.5}}
\put(0,-1){\line(1,0){17.5}} \put(0,1){\line(1,0){17.5}}
\put(0,50){\line(1,0){17.5}} \put(0,70){\line(1,0){17.5}}
\end{picture}}
\end{picture}}
\end{picture}
%
\\
%
\begin{picture}(200,120)(0,0)
\put(10,5){
\begin{picture}(120,70)(0,0)
\put(-10,40){$\langle V_{14} \rangle=$}
\put(70,0){
\begin{picture}(20,100)(0,0)
\put(0,7){$\Gamma^{qq}(p',p)$} \put(0,97){$\Gamma^{qq}(p,p')$}
\put(-9,37){$V(p\hspace{-.1cm}-\hspace{-.1cm}p')$}
\end{picture}}
\put(20,0){
\begin{picture}(30,100)(0,0)
\put(10,8){$\phi$} \put(20,-5){\line(0,1){30}}
\put(0,10){\line(4,3){20}} \put(0,10){\line(4,-3){20}}
\put(10,70){$\psi$} \put(20,45){\line(0,1){50}}
\put(0,70){\line(4,5){20}} \put(0,70){\line(4,-5){20}}
\end{picture}}
\put(147.5,0){
\begin{picture}(30,100)(0,0)
\put(3,8){$\phi$} \put(0,-5){\line(0,1){30}}
\put(20,10){\line(-4,3){20}} \put(20,10){\line(-4,-3){20}}
\put(3,70){$\psi$} \put(0,45){\line(0,1){50}}
\put(20,70){\line(-4,5){20}} \put(20,70){\line(-4,-5){20}}
\end{picture}}
\put(53,0){
\begin{picture}(20,100)(0,0)
\put(0,10){\oval(20,14)[l]} \put(-10,10){\vector(0,-1){2}}
\put(-5,8){$p'$} \put(0,60){\oval(20,14)[l]}
\put(-10,60){\vector(0,-1){2}} \put(-5,56){$k$}
\put(0,80){\oval(20,14)[l]} \put(-10,80){\vector(0,-1){2}}
\put(-5,78){$p$}
\end{picture}}
\put(135,0){
\begin{picture}(20,100)(0,0)
\put(0,10){\oval(20,14)[r]} \put(10,10){\vector(0,1){2}}
\put(0,8){$p'$} \put(0,60){\oval(20,14)[r]}
\put(10,60){\vector(0,1){2}} \put(0,56){$k$}
\put(0,80){\oval(20,14)[r]} \put(10,80){\vector(0,1){2}}
\put(0,78){$p$}
\end{picture}}
\put(40,0){
\begin{picture}(30,100)(0,0)
\put(0,-1){\line(1,0){30}} \put(0,1){\line(1,0){30}}
\put(0,20){\line(1,0){30}} \put(0,50){\line(1,0){30}}
\put(0,70){\line(1,0){30}} \put(0,90){\line(1,0){30}}
\end{picture}}
\put(117.5,0){
\begin{picture}(50,100)(0,0)
\put(0,-1){\line(1,0){30}} \put(0,1){\line(1,0){30}}
\put(0,20){\line(1,0){30}} \put(0,50){\line(1,0){30}}
\put(0,70){\line(1,0){30}} \put(0,90){\line(1,0){30}}
\end{picture}}
\put(70,0){
\begin{picture}(20,100)(0,0)
\put(0,-1){\line(1,0){30}} \put(0,1){\line(1,0){30}}
\put(0,20){\line(1,0){30}} \put(0,50){\line(1,0){30}}
\put(0,70){\line(1,0){30}} \put(0,90){\line(1,0){30}}
\multiput(0,21.5)(0,5){14}{\line(0,1){2}} \put(0,20){\circle*{2}}
\put(0,90){\circle*{2}}
\end{picture}}
\put(100,0){
\begin{picture}(20,100)(0,0)
\put(0,20){\line(1,4){17.5}} \put(0,90){\line(1,-4){17.5}}
\put(0,-1){\line(1,0){17.5}} \put(0,1){\line(1,0){17.5}}
\put(0,50){\line(1,0){17.5}} \put(0,70){\line(1,0){17.5}}
\end{picture}}
\end{picture}}
\end{picture}
%
\begin{picture}(200,120)(0,0)
\put(10,5){
\begin{picture}(120,70)(0,0)
\put(-10,40){$\langle V_{25} \rangle=$}
\put(20,0){
\begin{picture}(30,100)(0,0)
\put(10,8){$\phi$} \put(20,-5){\line(0,1){30}}
\put(0,10){\line(4,3){20}} \put(0,10){\line(4,-3){20}}
\put(10,70){$\psi$} \put(20,45){\line(0,1){50}}
\put(0,70){\line(4,5){20}} \put(0,70){\line(4,-5){20}}
\end{picture}}
\put(147.5,0){
\begin{picture}(30,100)(0,0)
\put(3,8){$\phi$} \put(0,-5){\line(0,1){30}}
\put(20,10){\line(-4,3){20}} \put(20,10){\line(-4,-3){20}}
\put(3,70){$\psi$} \put(0,45){\line(0,1){50}}
\put(20,70){\line(-4,5){20}} \put(20,70){\line(-4,-5){20}}
\end{picture}}
\put(40,0){
\begin{picture}(30,100)(0,0)
\put(0,-1){\line(1,0){30}} \put(0,1){\line(1,0){30}}
\put(0,20){\line(1,0){30}} \put(0,50){\line(1,0){30}}
\put(0,70){\line(1,0){30}} \put(0,90){\line(1,0){30}}
\end{picture}}
\put(117.5,0){
\begin{picture}(50,100)(0,0)
\put(0,-1){\line(1,0){30}} \put(0,1){\line(1,0){30}}
\put(0,20){\line(1,0){30}} \put(0,50){\line(1,0){30}}
\put(0,70){\line(1,0){30}} \put(0,90){\line(1,0){30}}
\end{picture}}
\put(70,0){
\begin{picture}(20,100)(0,0)
\put(0,-1){\line(1,0){30}} \put(0,1){\line(1,0){30}}
\put(0,20){\line(1,0){30}} \put(0,50){\line(1,0){30}}
\put(0,70){\line(1,0){30}} \put(0,90){\line(1,0){30}}
\multiput(0,1.5)(0,5){14}{\line(0,1){2}} \put(0,0){\circle*{3}}
\put(0,70){\circle*{2}}
\end{picture}}
\put(100,0){
\begin{picture}(20,100)(0,0)
\put(0,20){\line(1,4){17.5}} \put(0,90){\line(1,-4){17.5}}
\put(0,-1){\line(1,0){17.5}} \put(0,1){\line(1,0){17.5}}
\put(0,50){\line(1,0){17.5}} \put(0,70){\line(1,0){17.5}}
\end{picture}}
\end{picture}}
\end{picture}
%
\\
%
\begin{picture}(200,120)(0,0)
\put(10,5){
\begin{picture}(120,70)(0,0)
\put(-10,40){$\langle V_{23} \rangle=$}
\put(20,0){
\begin{picture}(30,100)(0,0)
\put(10,8){$\phi$} \put(20,-5){\line(0,1){30}}
\put(0,10){\line(4,3){20}} \put(0,10){\line(4,-3){20}}
\put(10,70){$\psi$} \put(20,45){\line(0,1){50}}
\put(0,70){\line(4,5){20}} \put(0,70){\line(4,-5){20}}
\end{picture}}
\put(147.5,0){
\begin{picture}(30,100)(0,0)
\put(3,8){$\phi$} \put(0,-5){\line(0,1){30}}
\put(20,10){\line(-4,3){20}} \put(20,10){\line(-4,-3){20}}
\put(3,70){$\psi$} \put(0,45){\line(0,1){50}}
\put(20,70){\line(-4,5){20}} \put(20,70){\line(-4,-5){20}}
\end{picture}}
\put(40,0){
\begin{picture}(30,100)(0,0)
\put(0,-1){\line(1,0){30}} \put(0,1){\line(1,0){30}}
\put(0,20){\line(1,0){30}} \put(0,50){\line(1,0){30}}
\put(0,70){\line(1,0){30}} \put(0,90){\line(1,0){30}}
\end{picture}}
\put(117.5,0){
\begin{picture}(50,100)(0,0)
\put(0,-1){\line(1,0){30}} \put(0,1){\line(1,0){30}}
\put(0,20){\line(1,0){30}} \put(0,50){\line(1,0){30}}
\put(0,70){\line(1,0){30}} \put(0,90){\line(1,0){30}}
\end{picture}}
\put(70,0){
\begin{picture}(20,100)(0,0)
\put(0,-1){\line(1,0){30}} \put(0,1){\line(1,0){30}}
\put(0,20){\line(1,0){30}} \put(0,50){\line(1,0){30}}
\put(0,70){\line(1,0){30}} \put(0,90){\line(1,0){30}}
\multiput(0,49)(0,5){5}{\line(0,1){2}} \put(0,50){\circle*{2}}
\put(0,70){\circle*{2}}
\end{picture}}
\put(100,0){
\begin{picture}(20,100)(0,0)
\put(0,20){\line(1,4){17.5}} \put(0,90){\line(1,-4){17.5}}
\put(0,-1){\line(1,0){17.5}} \put(0,1){\line(1,0){17.5}}
\put(0,50){\line(1,0){17.5}} \put(0,70){\line(1,0){17.5}}
\end{picture}}
\end{picture}}
\end{picture}
%
\begin{picture}(200,120)(0,0)
\put(10,5){
\begin{picture}(120,70)(0,0)
\put(-10,40){$\langle A_{15} \rangle=$}
\put(20,0){
\begin{picture}(30,100)(0,0)
\put(10,8){$\phi$} \put(20,-5){\line(0,1){30}}
\put(0,10){\line(4,3){20}} \put(0,10){\line(4,-3){20}}
\put(10,70){$\psi$} \put(20,45){\line(0,1){50}}
\put(0,70){\line(4,5){20}} \put(0,70){\line(4,-5){20}}
\end{picture}}
\put(147.5,0){
\begin{picture}(30,100)(0,0)
\put(3,8){$\phi$} \put(0,-5){\line(0,1){30}}
\put(20,10){\line(-4,3){20}} \put(20,10){\line(-4,-3){20}}
\put(3,70){$\psi$} \put(0,45){\line(0,1){50}}
\put(20,70){\line(-4,5){20}} \put(20,70){\line(-4,-5){20}}
\end{picture}}
\put(40,0){
\begin{picture}(30,100)(0,0)
\put(0,-1){\line(1,0){30.5}} \put(0,1){\line(1,0){29.5}}
\put(0,20){\line(1,0){30}} \put(0,50){\line(1,0){30}}
\put(0,70){\line(1,0){30}} \put(0,90){\line(1,0){30}}
\end{picture}}
\put(117.5,0){
\begin{picture}(50,100)(0,0)
\put(-.5,-1){\line(1,0){30.5}} \put(.5,1){\line(1,0){29.5}}
\put(0,20){\line(1,0){30}} \put(0,50){\line(1,0){30}}
\put(0,70){\line(1,0){30}} \put(0,90){\line(1,0){30}}
\end{picture}}
\put(70,0){
\begin{picture}(20,100)(0,0)
\put(.5,-1){\line(1,4){10}} \put(-.5,1){\line(1,4){10}}
\put(0,90){\line(1,-5){10}} \put(47,-1){\line(-1,4){10}}
\put(48,1){\line(-1,4){10}} \put(47.5,90){\line(-1,-5){10}}
\put(0,20){\line(1,0){47.5}} \put(0,50){\line(1,0){47.5}}
\put(0,70){\line(1,0){47.5}}
\multiput(10,40)(5,0){6}{\line(1,0){2}} \put(10,40){\circle*{3}}
\put(37.5,40){\circle*{3}}
\end{picture}}
\end{picture}}
\end{picture}
%
\caption{ Energy, kinetic and potential overlaps contributing to
meson-hadron scattering. The dot corresponds to the insertion of a
kinetic operator.} \label{RGM diagrams}
\end{figure}
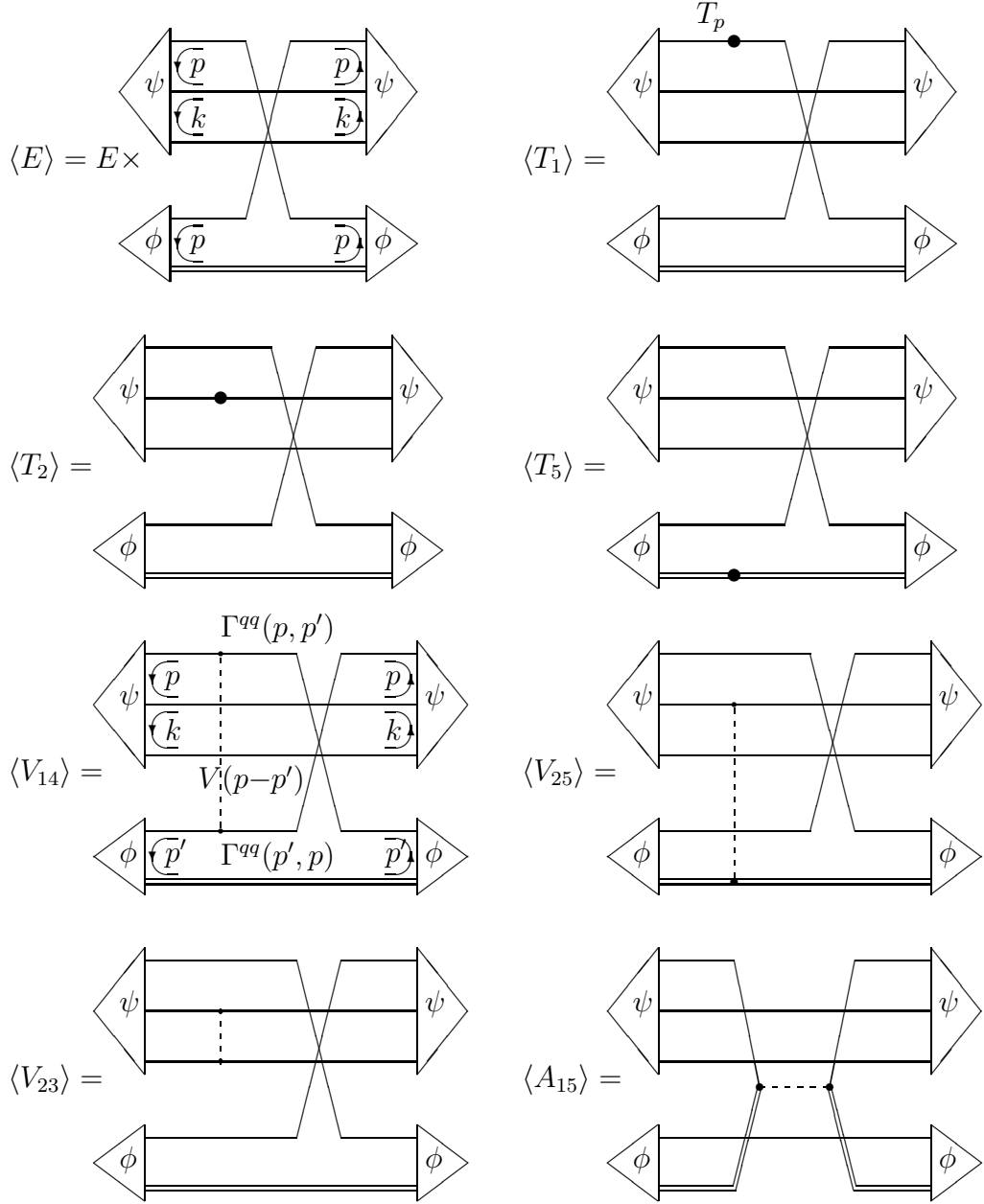

To compute the $T_{\pi-B}$ transition matrix we follow the
RGM to obtain the effective $\pi - B$
potential $\mathcal{O}^{\mbox{\tiny RGM}}_{\pi-B}$. To see how to
derive the RGM formalism for eNJL Hamiltonians see ref.
\cite{BicRib2,Bicudo,Bicudo4,Bicudo8,Bicudo9,Bicudo10},
where it was shown that the RGM equations for
hadronic scattering, being dynamical equations for coupled
channels of boundstates and resonances scattering, are simply
obtained by replacing in the S matrix Dyson equations the
correspondent solutions of the Bethe-Salpeter equations for each
of the intervening hadrons. We get
\begin{equation}  \label{eq:overlapkernel}
\mathcal{O}^{\mbox{\tiny RGM}}_{\pi-B}= \langle
3E-6T_1-6T_2-3T_5-3V_{23}+3V_{14}+6V_{25}+3A_{15} \rangle \ ,
\end{equation}
where the factors 3 and 6 are there to account for all the
possible permutations in the exchange of quarks and the minus
signs are produced by the Pauli principle. The existence of
negative energy spin amplitudes for the mesons complicates
considerably the RGM calculations. In order to take into account
this degree of freedom we will proceed in two steps. First we will
only consider those diagrams containing positive energy amplitudes
and later we extend this set of diagrams to those containing the
negative energy amplitudes. For the moment let us consider only
the positive energy amplitudes.

The different terms of Eq. (\ref{eq:overlapkernel}) are
represented by the diagrams in Fig. \ref{RGM diagrams}. For
instance diagrams $\langle E \rangle$ and $\langle V_{14} \rangle$
correspond to the integrals
\begin{eqnarray}
\langle E \rangle&=&\sum_{s_i} \int {\frac{d^3 p}{(2\pi)^3}}{\frac{d^3 k}{(2\pi)^3}}
\Psi_{s_1,s_2,s_3}^\dagger(\trD{p},\trD{k})\ \Psi_{s_4,s_2,s_3}(\trD{p},\trD{k})
\ \phi_{s_1,s_5}^\dagger(\trD{p})\ \phi_{s_4,s_5}(\trD{p}) \ ,\\
\langle V_{14} \rangle&=&
\sum_{s_i} \int{\frac{d^3 p}{(2\pi)^3}}{\frac{d^3 p'}{(2\pi)^3}}{\frac{d^3 k}{(2\pi)^3}}
\Psi_{s_1,s_2,s_3}^\dagger(\trD{p},\trD{k})\ \Psi_{s_7,s_2,s_3}(\trD{p},\trD{k})
\nonumber \\
&&
\hspace{1cm} \Gamma^{qq}_{s_1,s_6}(\trD{p},\trD{p'})\ V(\trD{p}-\trD{p'})
\ \Gamma^{qq}_{s_4,s_7}(\trD{p'},\trD{p})
\ \phi^\dagger_{s_4,s_5}(\trD{p'})\ \phi_{s_6,s_5}(\trD{p'}) \ ,
\label{O-contrib}
\end{eqnarray}
where $V(\trD{p}-\trD{p'})$ represents the Fourier transform of a
generic quark kernel $K(x-y)$ which we do not need to know. In Eq.
(\ref{O-contrib}) $\phi$ and $\Psi$ are the Salpeter amplitudes
for pions and baryons respectively.

Fig. \ref{RGM diagrams} includes all topologically possible diagrams,
except for diagrams which interaction can be included from the onset in
the energy of the external bound states. All diagrams, except for
$\langle A_{15} \rangle$, include a quark exchange because a simple kernel
insertion (which goes with
${ \vec \lambda \over 2} \cdot { \vec \lambda \over 2}$ in color space)
has zero matrix element between color singlets.

The diagrams $\langle V_{14} \rangle$ and $\langle V_{25} \rangle$ are the
usual RGM diagrams yielding for most reactions a repulsive force. This is
the case of Nucleon-Nucleon elastic scattering where they are shown to
provide a substantial repulsion
\cite{Ribeiro1,Ribeiro3}.
The same happens with
exotic meson-meson reactions. Retrospectively we can say that the success
of the present calculation constitutes an extension of these results.

The diagram $\langle A_{15} \rangle$ contains quark-antiquark
annihilation amplitudes and for most cases it provides for
attraction
\cite{Deus}.

Then $a_{\pi-B}$ is given by
\begin{equation}
\label{exact}
a_{\pi-B}=- \mu_{red} \mathcal{O}^{\mbox{\tiny RGM}}_{\pi-B} /(2 \pi+2 \mu_{red}
\mathcal{O}^{\mbox{\tiny RGM}}_{\pi-B} N^{2/3}).
\end{equation}
In Eq. (\ref{exact}) $\mu_{red}$ is the reduced mass of the system
$\pi-B$,
\begin{equation}
\mu_{red}=M_\pi-M_\pi^2/M_B+\cdots \ ,
\end{equation}
and the simple Born approximation already provides the leading
order in $M_\pi$
\begin{equation} \label{eq:aBorn}
a_{\pi-B}=-M_\pi \mathcal{O}^{\mbox{\tiny RGM}}_{\pi-B} /(2 \pi) \
.
\end{equation}

In order to calculate $\mathcal{O}^{\mbox{\tiny RGM}}_{\pi-B}$ we
need first to know the Salpeter solutions for the asymptotic pions
and baryons which will then act as wave-functions for the ordinary
RGM equations. As we have already stated this recipe is equivalent
to solve the complete Dyson series for $\pi-B$ scattering for
contact diagrams. For a detailed proof of this equivalence see
ref.
\cite{BicRib2,Bicudo,Bicudo4,Bicudo8,Bicudo9,Bicudo10}.

\section{Salpeter equations for mesons and baryons}
\label{bound state}

In this section we give the Salpeter amplitudes for mesons and
$B$s. Since we are considering s-wave hadrons the angular momentum
is trivial, and the spin is expected to dominate dynamics. The
independent spin degrees of freedom correspond to a spin singlet
and a spin triplet. For instance, in the $\Delta$ the diquarks are
in a spin triplet state whereas in the $N$ they have a mixed spin
structure $(F_{\mbox{\tiny{spin}}}F_{\mbox{\tiny{flavor}}}+
D_{\mbox{\tiny{spin}}}D_{\mbox{\tiny{flavor}}})/\sqrt{2}$, where
the $F$ component is a singlet and the $D$ component a triplet. In
the meson sector the spin structure is clearer. The $\pi$ meson is
a spin singlet and the $\rho$ meson is a spin triplet. Besides
spin structure we have another degree of freedom known as the
energy-spin which takes into account the positive ad negative
energy Salpeter amplitudes. These negative energy amplitudes play
a major role in the Goldstone nature of the pion, being much less
relevant for the $\rho$ case. In the present work these mesonic
negative energy amplitudes will be fully taken into account. There
are no negative energy amplitudes for the $B$s
\cite{Bicudo7}.

\begin{figure}
%
\begin{picture}(120,90)(0,0)
\put(20,45){
\begin{picture}(120,70)(0,0)
\put(-15,8){$[E-2T_p]\times$}
\put(70,0){
\begin{picture}(30,100)(0,0)
\put(2,5){$\phi\hspace{-.05cm}^+$}
\put(0,-5){\line(0,1){30}}
\put(20,10){\line(-4,3){20}}
\put(20,10){\line(-4,-3){20}}
\end{picture}}
\put(57.5,0){
\begin{picture}(20,100)(0,0)
\put(0,10){\oval(20,14)[r]}
\put(10,10){\vector(0,1){2}}
\put(0,8){$p$}
\end{picture}}
\put(50,0){
\begin{picture}(50,100)(0,0)
\put(0,-1){\line(1,0){20}}
\put(0,1){\line(1,0){20}}
\put(0,20){\line(1,0){20}}
\end{picture}}
\end{picture}}
\end{picture}
%
%
\begin{picture}(120,90)(0,0)
\put(5,45){
\begin{picture}(120,70)(0,0)
\put(0,0){$=$}
\put(10,0){
\begin{picture}(20,100)(0,0)
\put(10,-13){$\Gamma^{\bar q \bar q}(p,p')$}
\put(10,27){$\Gamma^{qq}(p,p')$}
\put(5,7){$V(p\hspace{-.1cm}-\hspace{-.1cm}p')$}
\end{picture}}
\put(80,0){
\begin{picture}(30,100)(0,0)
\put(2,5){$\phi\hspace{-.05cm}^+$}
\put(0,-5){\line(0,1){30}}
\put(20,10){\line(-4,3){20}}
\put(20,10){\line(-4,-3){20}}
\end{picture}}
\put(40,0){
\begin{picture}(50,100)(0,0)
\put(0,-1){\line(1,0){20}}
\put(0,1){\line(1,0){20}}
\put(0,20){\line(1,0){20}}
\end{picture}}
\put(60,0){
\begin{picture}(20,100)(0,0)
\put(0,-1){\line(1,0){20}}
\put(0,1){\line(1,0){20}}
\put(0,20){\line(1,0){20}}
\multiput(0,1.5)(0,5){4}{\line(0,1){2}}
\put(0,0){\circle*{3}}
\put(0,20){\circle*{2}}
\end{picture}}
\end{picture}}
\end{picture}
%
\begin{picture}(100,90)(0,0)
\put(15,5){
\begin{picture}(120,70)(0,0)
\put(-15,40){$+$}
\put(0,0){
\begin{picture}(30,100)(0,0)
\put(7,8){$\phi\hspace{-.05cm}^-$}
\put(20,-5){\line(0,1){30}}
\put(0,10){\line(4,3){20}}
\put(0,10){\line(4,-3){20}}
\end{picture}}
\put(7.5,40){
\begin{picture}(20,100)(0,0)
\put(0,10){\oval(20,14)[r]}
\put(10,10){\vector(0,1){2}}
\put(0,8){$p$}
\end{picture}}
\put(33,0){
\begin{picture}(20,100)(0,0)
\put(0,10){\oval(20,14)[l]}
\put(-10,10){\vector(0,-1){2}}
\put(-5,8){$p'$}
\end{picture}}
\put(0,0){
\begin{picture}(30,100)(0,0)
\put(20,-1){\line(1,0){20.5}}
\put(20,1){\line(1,0){20}}
\put(20,20){\line(1,0){20}}
\put(0,39){\line(1,0){40}}
\put(0,41){\line(1,0){40.5}}
\put(0,60){\line(1,0){40}}
\end{picture}}
\put(50,30){
\begin{picture}(20,100)(0,0)
\put(0,0){\line(-1,-1){10}}
\put(20.5,-1){\line(-1,-1){30}}
\put(20,1){\line(-1,-1){30}}
\put(0,-1){\line(-1,1){10}}
\put(0.5,1){\line(-1,1){10}}
\put(20,0){\line(-1,1){30}}
\multiput(1.5,0)(5,0){4}{\line(1,0){2}}
\put(0,0){\circle*{3}}
\put(20,0){\circle*{3}}
\end{picture}}
\end{picture}}
\end{picture}
%
\\
%
\begin{picture}(120,70)(0,0)
\put(20,5){
\begin{picture}(120,70)(0,0)
\put(-20,8){$[-E-2T_p]\times$}
\put(50,0){
\begin{picture}(30,100)(0,0)
\put(7,8){$\phi\hspace{-.05cm}^-$}
\put(20,-5){\line(0,1){30}}
\put(0,10){\line(4,3){20}}
\put(0,10){\line(4,-3){20}}
\end{picture}}
\put(50,0){
\begin{picture}(30,100)(0,0)
\put(20,-1){\line(1,0){20}}
\put(20,1){\line(1,0){20}}
\put(20,20){\line(1,0){20}}
\end{picture}}
\put(83,0){
\begin{picture}(20,100)(0,0)
\put(0,10){\oval(20,14)[l]}
\put(-10,10){\vector(0,-1){2}}
\put(-5,8){$p$}
\end{picture}}
\end{picture}}
\end{picture}
%
\begin{picture}(120,70)(0,0)
\put(0,5){
\begin{picture}(120,70)(0,0)
\put(0,5){$=$}
\put(80,0){
\begin{picture}(30,100)(0,0)
\put(20,-1){\line(-1,0){40.5}}
\put(20,1){\line(-1,0){40}}
\put(20,20){\line(-1,0){40}}
\put(0,39){\line(-1,0){20}}
\put(0,41){\line(-1,0){20.5}}
\put(0,60){\line(-1,0){20}}
\end{picture}}
\put(50,30){
\begin{picture}(20,100)(0,0)
\put(0,0){\line(1,-1){10}}
\put(-20.5,-1){\line(1,-1){30}}
\put(-20,1){\line(1,-1){30}}
\put(0,-1){\line(1,1){10}}
\put(-0.5,1){\line(1,1){10}}
\put(-20,0){\line(1,1){30}}
\multiput(-18.5,0)(5,0){4}{\line(1,0){2}}
\put(0,0){\circle*{3}}
\put(-20,0){\circle*{3}}
\end{picture}}
\put(80,40){
\begin{picture}(30,100)(0,0)
\put(2,5){$\phi\hspace{-.05cm}^+$}
\put(0,-5){\line(0,1){30}}
\put(20,10){\line(-4,3){20}}
\put(20,10){\line(-4,-3){20}}
\end{picture}}
\put(10,0){
\begin{picture}(20,100)(0,0)
\put(0,27){$\Gamma^{\bar q q}$}
\put(48,26){$\Gamma^{\bar q q}$}
\put(33,33){$V$}
\end{picture}}
\end{picture}}
\end{picture}
%
\begin{picture}(90,70)(0,0)
\put(0,5){
\begin{picture}(120,70)(0,0)
\put(0,0){$+$}
\put(10,0){
\begin{picture}(30,100)(0,0)
\put(7,8){$\phi\hspace{-.05cm}^-$}
\put(20,-5){\line(0,1){30}}
\put(0,10){\line(4,3){20}}
\put(0,10){\line(4,-3){20}}
\end{picture}}
\put(30,0){
\begin{picture}(50,100)(0,0)
\put(0,-1){\line(1,0){20}}
\put(0,1){\line(1,0){20}}
\put(0,20){\line(1,0){20}}
\end{picture}}
\put(50,0){
\begin{picture}(20,100)(0,0)
\put(0,-1){\line(1,0){20}}
\put(0,1){\line(1,0){20}}
\put(0,20){\line(1,0){20}}
\multiput(0,1.5)(0,5){4}{\line(0,1){2}}
\put(0,0){\circle*{3}}
\put(0,20){\circle*{2}}
\end{picture}}
\put(43,0){
\begin{picture}(20,100)(0,0)
\put(0,10){\oval(20,14)[l]}
\put(-10,10){\vector(0,-1){2}}
\put(-5,8){$p'$}
\end{picture}}
\put(63,0){
\begin{picture}(20,100)(0,0)
\put(0,10){\oval(20,14)[l]}
\put(-10,10){\vector(0,-1){2}}
\put(-5,8){$p$}
\end{picture}}
\end{picture}}
\end{picture}
%
%
\caption{Salpeter coupled equations for a quark and an antiquark in a meson.
In most mesons the negative energy component $\phi^-$ is negligible which
simplifies the equation to a single Schr\"odinger equation}
\label{Salpeter diagrams}
\end{figure}

The bound state equations for hadrons are essentially
quark-antiquark Salpeter or RPA equation in the case of mesons,
see Fig. \ref{Salpeter diagrams}, and  Schr\"odinger or TDA
equation for three quarks case of baryons, see Fig.
\ref{Schrodinger diagrams}. In Fig. \ref{Salpeter diagrams} the
mesonic Salpeter equation consists in fact of two coupled
equations, coupling positive to negative energy amplitudes. As an
example, we can use the vertices definitions of Table
\ref{vertices} to write one of these equations (the topmost) as
\begin{equation} \label{mesonbseq}
\begin{split}
H \phi^+_{s_1,s_2}(p)&= [T_p+T_{-p}]\phi^+_{s_1,s_2}(p) +\\
& \hspace{.5cm}
+\sum_{s_3,s_4} \int { d^3 p' \over (2\pi)^3 } \Gamma^{qq}_{s_1,s_3}(\trD{p},\trD{p}')
V(\trD{p}-\trD{p}') \Gamma^{\bar q \bar q}_{s_2,s_4}(\trD{p}',\trD{p}) \phi^+_{s_3,s_4}(p') + \\
& \hspace{.5cm}
+\sum_{s_3,s_4} \int { d^3 p' \over (2\pi)^3 }
\Gamma^{q \bar q}_{s_1,s_3}(\trD{p},\trD{p}') V(\trD{p}-\trD{p}')
\Gamma^{\bar q q}_{s_2,s_4}(\trD{p}',\trD{p}) \phi^-_{s_3,s_4}(p') \\
&= E \phi^+_{s_1,s_2}(p) \ .
\end{split}
\end{equation}
In particular Eq. (\ref{mesonbseq}) reduces to the Schr\"odinger
equation for mesons when the coupling to the negative energy
$\phi^-$ wave-function is negligible.

\begin{figure}
%
\begin{picture}(110,70)(0,0)
\put(0,5){
\begin{picture}(120,70)(0,0)
\put(-5,27){$[ E-T_p$}
\put(-5,10){$\hspace{.1cm}-T_{k-p}-T_k ]$}
\put(60,0){
\begin{picture}(30,100)(0,0)
\put(0,0){\line(1,0){20}}
\put(0,20){\line(1,0){20}}
\put(0,40){\line(1,0){20}}
\end{picture}}
\put(67.5,0){
\begin{picture}(20,100)(0,0)
\put(0,10){\oval(20,14)[r]}
\put(10,10){\vector(0,1){2}}
\put(0,6){$k$}
\put(0,30){\oval(20,14)[r]}
\put(10,30){\vector(0,1){2}}
\put(0,28){$p$}
\end{picture}}
\put(80,0){
\begin{picture}(30,100)(0,0)
\put(3,20){$\psi$}
\put(0,-5){\line(0,1){50}}
\put(20,20){\line(-4,5){20}}
\put(20,20){\line(-4,-5){20}}
\end{picture}}
\end{picture}}
\end{picture}
%
\begin{picture}(115,70)(0,0)
\put(20,5){
\begin{picture}(120,70)(0,0)
\put(-20,20){$=$}
\put(10,0){
\begin{picture}(30,100)(0,0)
\put(20,0){\line(1,0){40}}
\put(20,20){\line(1,0){40}}
\put(20,40){\line(1,0){40}}
\end{picture}}
\put(10,0){
\begin{picture}(20,100)(0,0)
\put(-15,7){$\Gamma^{qq}(\hspace{-.05cm}k\hspace{-.1cm}-
\hspace{-.1cm}p,\hspace{-.05cm}k\hspace{-.1cm}-\hspace{-.1cm}p')$}
\put(-3,27){$V(p\hspace{-.1cm}-\hspace{-.1cm}p')$}
\put(5,47){$\Gamma^{qq}(p,p')$}
\end{picture}}
\put(50,0){
\begin{picture}(20,100)(0,0)
\multiput(0,21.5)(0,5){4}{\line(0,1){2}}
\put(0,20){\circle*{2}}
\put(0,40){\circle*{2}}
\end{picture}}
\put(70,0){
\begin{picture}(30,100)(0,0)
\put(3,20){$\psi$}
\put(0,-5){\line(0,1){50}}
\put(20,20){\line(-4,5){20}}
\put(20,20){\line(-4,-5){20}}
\end{picture}}
\end{picture}}
\end{picture}
%
\begin{picture}(80,70)(0,0)
\put(0,5){
\begin{picture}(120,70)(0,0)
\put(0,20){$+$}
\put(10,0){
\begin{picture}(30,100)(0,0)
\put(0,0){\line(1,0){40}}
\put(0,20){\line(1,0){40}}
\put(0,40){\line(1,0){40}}
\end{picture}}
\put(30,0){
\begin{picture}(20,100)(0,0)
\multiput(0,1.5)(0,5){4}{\line(0,1){2}}
\put(0,0){\circle*{2}}
\put(0,20){\circle*{2}}
\end{picture}}
\put(37.5,0){
\begin{picture}(20,100)(0,0)
\put(0,10){\oval(20,14)[r]}
\put(10,10){\vector(0,1){2}}
\put(-2,6){$k'$}
\put(-22,10){\oval(20,14)[r]}
\put(-12,10){\vector(0,1){2}}
\put(-22,6){$k$}
\put(0,30){\oval(20,14)[r]}
\put(10,30){\vector(0,1){2}}
\put(0,28){$p$}
\end{picture}}
\put(50,0){
\begin{picture}(30,100)(0,0)
\put(3,20){$\psi$}
\put(0,-5){\line(0,1){50}}
\put(20,20){\line(-4,5){20}}
\put(20,20){\line(-4,-5){20}}
\end{picture}}
\end{picture}}
\end{picture}
%
\begin{picture}(80,70)(0,0)
\put(0,5){
\begin{picture}(120,70)(0,0)
\put(0,20){$+$}
\put(10,0){
\begin{picture}(30,100)(0,0)
\put(0,0){\line(1,0){40}}
\put(0,20){\line(1,0){40}}
\put(0,40){\line(1,0){40}}
\end{picture}}
\put(30,0){
\begin{picture}(20,100)(0,0)
\multiput(0,1.5)(0,5){8}{\line(0,1){2}}
\put(0,0){\circle*{2}}
\put(0,40){\circle*{2}}
\end{picture}}
\put(50,0){
\begin{picture}(30,100)(0,0)
\put(3,20){$\psi$}
\put(0,-5){\line(0,1){50}}
\put(20,20){\line(-4,5){20}}
\put(20,20){\line(-4,-5){20}}
\end{picture}}
\end{picture}}
\end{picture}
%
\caption{Schr\"odinger equation for 3 quarks in a baryon}
\label{Schrodinger diagrams}
\end{figure}

The next step consists in working out the spin$\times$flavour$\times$colour
traces. These traces are worked out in appendix A. Doing this will lead us,
in the case of mesons, to the following two coupled channel equations,
\begin{equation}
H_m(\trD{p})\;\raisebox{-6pt} {$
 \left[
 \begin{array}{c}
  | \phi_m^+ \rangle \\
  | \phi_m^- \rangle
 \end{array}\right]$}\;
= \int\frac{d^3q}{(2\pi)^3}
\mathcal{H}(\trD{p},\trD{q})
\raisebox{-6pt}{$
 \left[
 \begin{array}{c}
  | \phi_m^+ \rangle \\
  | \phi_m^- \rangle
 \end{array}\right]$}= M_m \sigma^3
\raisebox{-6pt}{$
 \left[
 \begin{array}{c}
  | \phi_m^+ \rangle \\
  | \phi_m^- \rangle
 \end{array}\right]$} \ ,
\end{equation}
with
\begin{equation}
\mathcal{H}(\trD{p},\trD{q})=
\raisebox{-6pt}{$
 \left[
 \begin{array}{cc}
  2T_p\,\delta(\trD{p}-\trD{q})+V_m^{++}(\trD{p},\trD{q}) & V_m^{+-}(\trD{p},\trD{q}) \\
  V_m^{-+}(\trD{p},\trD{q}) & 2T_p\,\delta(\trD{p}-\trD{q})+V_m^{--}(\trD{p},\trD{q})
 \end{array}\right]$} \ ,
\end{equation}
where $V_m^{++}$, $V_m^{+-}$, $V_m^{-+}$ and $V_m^{--}$ contain
the aforementioned spin traces.

\subsection{The pion case}
First we work out the spin traces for the pion case. We have,
\begin{equation}   \label{bspi}
\begin{split}
 V_\pi^{++}(\trD{p},\trD{p}') &= \frac{4}{3}\ Tr\{\Sigma^{\pi}\cdot
\Gamma^{\bar q \bar q}_{p,p'}\cdot{\Sigma^\pi}^\dag\cdot\Gamma^{qq}_{p',p}\}
V(\trD{p}-\trD{p}') \\
&= -\frac{2}{3}\ Tr\{\Gamma^{qq}_{p,p'}\Gamma^{qq}_{p',p}\}V(\trD{p}-\trD{p}')\ , \\
V_\pi^{+-}(\trD{p},\trD{p}') &= \frac{4}{3}\ Tr\{\Sigma^\pi\cdot
\Gamma^{\bar q q }_{p,p'}\cdot\Sigma^\pi\cdot \Gamma^{\bar q q}_{p',p}\}
V(\trD{p}-\trD{p}') \\
&= -\frac{2}{3}\ Tr\{\Gamma^{q \bar q}_{p,p'}\Gamma^{\bar q q}_{p',p}\}
V(\trD{p}-\trD{p}')\ .
\end{split}
\end{equation}
where $\Sigma^\pi$ stands for the pion spin wave function defined
in Appendix \ref{traces} along with spin wave functions for other
relevant hadrons. Notice that $V_\pi^{+-}$ acts as a potential
coupling the negative energy wave-function $\phi^-$ with the
positive energy wave-function $\phi^+$. $V_\pi^{++}$ stands as the
usual potential binding the quark and the antiquark in the
positive energy wave-function $\phi^+$. The color factor 4/3 is
included in the $V$'s definition for simplicity. The relations
\begin{equation} \label{resultmapa}
\Gamma^{qq}=(i\sigma_2)\Gamma^{\bar q \bar q}(i\sigma_2) \ ,
\hspace{2cm}
\Gamma^{q \bar q}=-(i\sigma_2)\Gamma^{\bar q q}(i\sigma_2) \ ,
\end{equation}
were used. They hold for both the Dirac vertices $\gamma^0$ and $\vec \gamma$.

The pion occupies a singular position, insofar it is the only
bound state where  $\phi^-_\pi$ is not negligible. In particular
$|\phi^-_\pi|$ is just slightly smaller than $|\phi^+_\pi|$
because the $\pi$ is nearly a Goldstone boson. The Salpeter pion
wave functions are
\begin{equation}
\phi^\pm(\trD{k})=\frac{S(\trD{k})}{a}\pm a\,\xi(\trD{k})
\mbox{\hspace{1cm}with\hspace{.5cm}} a=\sqrt{\frac{2 M_\pi
}{3}}f_\pi\ , \label{wavefunction}
\end{equation}
where $f_\pi$ is the well known pion decay constant, $S(k)$
determines the extent of dynamical chiral symmetry breaking and
$\xi(k)$ is related to normalization of the wave function. The
Salpeter wave functions are normalized as follows,
\begin{equation}  \label{eq:norm}
\begin{split}
&\left[
\begin{array}{cc}
\langle \phi^+ |  & \langle \phi^- |
\end{array}
\right] \sigma^3
\raisebox{-6pt}{$
 \left[
 \begin{array}{c}
|  \phi^+ \rangle \\
|  \phi^- \rangle
 \end{array}\right]$}=
\int\frac{d^3k}{(2\pi)^3}(\phi^+(\trD{k})^2-\phi^-(\trD{k})^2)= \\
&= 4\int\frac{d^3k}{(2\pi)^3}S(\trD{k})\xi(\trD{k}) =
\frac{2}{a}\int\frac{d^3k}{(2\pi)^3}S(\trD{k})(\phi^+(\trD{k})-\phi^+(\trD{k}))
=1 \ .
\end{split}
\end{equation}

\subsection{The rho case}
For the $\rho$ we can follow exactly the same steps as in the pion
case to get
\begin{equation}  \label{bsrho}
V_\rho^{++}(\trD{p},\trD{p}') =
-\frac{2}{9} Tr\{\sigma^k\Gamma^{qq}_{p,p'}\sigma^k \Gamma^{qq}_{p',p}\}
\delta_{s s'} V(\trD{p}-\trD{p}')\ .
\end{equation}
To obtain Eq. (\ref{bsrho}) the following identity
\begin{equation}
\int\frac{d^3p}{(2\pi)^3}\frac{d^3p'}{(2\pi)^3}
Tr\{\sigma^i\Gamma^{qq}_{p,p'}\sigma^j \Gamma^{qq}_{p',p}\} =
\int\frac{d^3p}{(2\pi)^3}\frac{d^3p'}{(2\pi)^3}
\frac{1}{3} Tr\{\sigma^k\Gamma^{qq}_{p,p'}\sigma^k \Gamma^{qq}_{p',p}\} \delta^{ij}
\ ,
\end{equation}
must be used.

The potential $V_\rho^{+-}(\trD{p},\trD{p}')$ can also be computed and it
turns out that it is exactly equal to $-1/3$ of $V_\pi^{+-}(\trD{p},\trD{p}')$.
It provides a negligible coupling of $\phi_\rho^-$ to $\phi_\rho^+$ and
this is the technical reason why the $\rho$ is essentially a Schr\"odinger
bound state.

\subsection{The baryon case}
In the case of the baryons we have no negative energy amplitude (there is no
way with a two body kernel of simultaneously reversing three quark lines)
and we need only to consider the $V_B^{++}$. Following the same steps used in
deriving Eqs. (\ref{bspi}) and (\ref{bsrho}) we get,
\begin{equation} \label{Vsimpl1}
\begin{split}
V_{N^F}^{++}(\trD{p},\trD{p}') & = 3\frac{2}{3}\
Tr\{\Sigma^{N^F}_s\cdot\Gamma^{qq}_{p-k,p'-k}\cdot{\Sigma^{N^F}_{s'}}^\dag\cdot
\Gamma^{qq}_{p,p'}\}\ V(\trD{p}-\trD{p}') \\
& = -2\frac{1}{2}\ Tr\{\Gamma^{qq}_{p'-k,p-k}\Gamma^{qq}_{p,p'}\}\
\delta_{ss'}V(\trD{p}-\trD{p}') \\
& = - Tr\{\Gamma^{qq}_{p'-k,p-k}\Gamma^{qq}_{p,p'}\}\
\delta_{ss'}V(\trD{p}-\trD{p}'),
\end{split}
\end{equation}
\begin{equation} \label{Vsimpl2}
\begin{split}
V_{N^D}^{++}(\trD{p},\trD{p}') & = 3\frac{2}{3}\
Tr\{\Sigma^{N^D}_s\cdot\Gamma^{qq}_{p-k,p'-k}\cdot {\Sigma^{N^D}_{s'}}^\dag
\cdot\Gamma^{qq}_{p,p'}\}\ V(\trD{p}-\trD{p}') \\
& = -2\frac{1}{6}\ Tr\{\sigma^i\Gamma^{qq}_{p'-k,p-k}\sigma^j
\Gamma^{qq}_{p,p'}\}\ \sigma^i_{sc}\sigma^j_{cs'}V(\trD{p}-\trD{p}') \\
& = -\frac{1}{3}\ Tr\{\sigma^k\Gamma^{qq}_{p'-k,p-k}\sigma^k
\Gamma^{qq}_{p,p'}\}\ \delta_{s s'}V(\trD{p}-\trD{p}'),
\end{split}
\end{equation}
\begin{equation} \label{Vsimpl3}
\begin{split}
V_{\Delta}^{++}(\trD{p},\trD{p}') & = 3\frac{2}{3}\
Tr\{\Sigma^\Delta_s\cdot\Gamma^{qq}_{p-k,p'-k}\cdot
{\Sigma^\Delta_{s'}}^\dag\cdot\Gamma^{qq}_{p,p'}\}\ V(\trD{p}-\trD{p}') \\
& = -3\frac{1}{3}\ Tr\{\sigma^i\Gamma^{qq}_{p'-k,p-k}\sigma^j\Gamma^{qq}_{p,p'}\}
\ {w^i_{sc}}^\dag w^j_{cs'}V(\trD{p}-\trD{p}') \\
& = -\frac{1}{3}\
Tr\{\sigma^k\Gamma^{qq}_{p'-k,p-k}\sigma^k\Gamma^{qq}_{p,p'}\}
\delta_{s s'}V(\trD{p}-\trD{p}').
\end{split}
\end{equation}
Along with the color factors we include the factor three that
accounts for the number of diagrams. We checked numerically, for
various microscopic kernels (like for example the $r^2$ harmonic
kernel) that, under integration in $k$, $V_{N^F}^{++}$,
$V_{N^D}^{++}$ and $V_{\Delta}^{++}$ are, to a good approximation,
given by,
\begin{equation} \label{Vapprox}
\begin{split}
V_{N^F}^{++}(\trD{p},\trD{p}')
& \simeq \frac{3}{2}V_\pi^{++}(\trD{p},\trD{p}')\delta_{ss'}, \\
V_{N^D}^{++}(\trD{p},\trD{p}') & \simeq
\frac{3}{2}V_\rho^{++}(\trD{p},\trD{p}')\delta_{ss'}, \\
V_{\Delta}^{++}(\trD{p},\trD{p}') & \simeq
\frac{3}{2}V_\rho^{++}(\trD{p},\trD{p}')\delta_{ss'},
\end{split}
\end{equation}
For the various quark kernels, the errors for the $N$ and $\Delta$
masses when using Eqs. (\ref{Vapprox}) instead of Eqs.
(\ref{Vsimpl1}), (\ref{Vsimpl2}) and (\ref{Vsimpl3}), were always
below $7\%$.

Having the spin$\times$flavor nucleon structure in mind -- see
Appendix \ref{traces} -- we get,
\begin{eqnarray}
V^{++}_N &=& \frac{3}{4}(V^{++}_\pi+V^{++}_\rho)\delta_{ss'}\ ,\nonumber\\
V^{++}_\Delta &=& \frac{3}{2}V^{++}_\rho\delta_{ss'} \label{eq:Vpot} \ .
\end{eqnarray}

In this case the baryon Salpeter amplitude will depend only on the
diquark inner momentum and the baryon Salpeter amplitude is given
by,
\begin{equation} \label{Bapprox}
\Psi(\trD{p})\simeq S(p)/\mathcal{N}
\end{equation}
where $\mathcal{N}$ is the baryon wave function normalization
which is essentially the same for both $N$ and $\Delta$.

\subsection{Summary}
The results of the previous subsections can be summarized as
follows,
\begin{equation}
\begin{split}
&H_\pi\;\raisebox{-6pt} {$
 \left[
 \begin{array}{c}
  | \phi_\pi^+ \rangle \\
  | \phi_\pi^- \rangle
 \end{array}\right]$}\; =\;
\int\frac{d^3 q}{(2\pi)^3}\mathcal{H}_\pi(\trD{p},\trD{q})
\raisebox{-6pt}{$
 \left[
 \begin{array}{c}
  | \phi_\pi^+ \rangle \\
  | \phi_\pi^- \rangle
 \end{array}\right]$}= M_\pi \sigma^3
\raisebox{-6pt}{$
 \left[
 \begin{array}{c}
  | \phi_\pi^+ \rangle \\
  | \phi_\pi^- \rangle
 \end{array}\right]$};  \\
&H_\rho | \phi_\rho \rangle = \int\frac{d^3
q}{(2\pi)^3}\mathcal{H}_\rho(\trD{p},\trD{q}) | \phi_\rho \rangle
= M_\rho | \phi_\rho \rangle; \\
&\mathcal{H}_\rho(\trD{p},\trD{q})=\left[ 2 T_p\,
\delta(\trD{p}-\trD{q})
+ V^{++}_\rho(\trD{p},\trD{q}) \right]; \\
&H_N | \Psi_N \rangle = \int\frac{d^3 q}{(2\pi)^3}
\mathcal{H}_N(\trD{p},\trD{q}) | \Psi_N \rangle = M_N
| \Psi_N \rangle; \\
&\mathcal{H}_N(\trD{p},\trD{q})=\frac{3}{2} \left[ 2 T_p\,
\delta(\trD{p}-\trD{q}) +
\frac{V^{++}_\pi(\trD{p},\trD{q})+V^{++}_\rho(\trD{p},\trD{q})}{2}
\right]; \\
&H_\Delta  | \Psi_\Delta  \rangle = \int\frac{d^3 q}{(2\pi)^3}
\mathcal{H}_\Delta(\trD{p},\trD{q}) | \Psi_\Delta \rangle
=M_\Delta | \Psi_\Delta  \rangle; \\
&\mathcal{H}_\Delta(\trD{p},\trD{q})=\frac{3}{2} \left[ 2 T_p\,
\delta(\trD{p}-\trD{q}) + V^{++}_\rho(\trD{p},\trD{q}) \right] \ .
\label{compact bound state}
\end{split}
\end{equation}
Eq. (\ref{compact bound state}) predicts that $M_\Delta = (3/2)
M_\rho$. This is a natural relation in the quark model and it is
experimentally correct within $7\%$ discrepancy. This suggests
that neglecting both the negative energy wave-function
$\phi^-_\rho$ and the diquark recoil in the baryons are not
unreasonable approximations for low energy scattering (scattering
lengths).

\section{Assembling the $\pi-B$ RGM overlaps
$\mathcal{O}^{\mbox{\tiny RGM}}_{\pi-B}$} \label{assembling}

Before computing the flavor contribution to the RGM overlaps of
Fig. \ref{RGM diagrams}, we have to extend the class of diagrams
in Fig. \ref{RGM diagrams} including the diagrams that couple to
the negative energy wave function $\phi^- _\pi$ of the external
pions. The negative energy wave function is defined in Fig.
\ref{Salpeter diagrams} and in Eqs. (\ref{mesonbseq}) and
(\ref{wavefunction}).

\begin{figure}
%
\begin{picture}(180,120)(0,0)
\put(0,5){
\begin{picture}(120,70)(0,0)
\put(0,40){$$}
\put(20,0){
\begin{picture}(30,100)(0,0)
\put(7,7){$\phi\hspace{-.05cm}^+$} \put(-20,7){out}
\put(20,-5){\line(0,1){30}} \put(0,10){\line(4,3){20}}
\put(0,10){\line(4,-3){20}} \put(10,70){$\psi$} \put(-20,70){out}
\put(20,45){\line(0,1){50}} \put(0,70){\line(4,5){20}}
\put(0,70){\line(4,-5){20}}
\end{picture}}
\put(120,0){
\begin{picture}(30,100)(0,0)
\put(1,6){$\phi\hspace{-.05cm}^+$} \put(23,7){in}
\put(0,-5){\line(0,1){30}} \put(20,10){\line(-4,3){20}}
\put(20,10){\line(-4,-3){20}} \put(3,70){$\psi$} \put(23,70){in}
\put(0,45){\line(0,1){50}} \put(20,70){\line(-4,5){20}}
\put(20,70){\line(-4,-5){20}}
\end{picture}}
\put(40,0){
\begin{picture}(30,100)(0,0)
\put(0,-1){\line(1,0){30}} \put(0,1){\line(1,0){30}}
\put(0,20){\line(1,0){30}} \put(0,50){\line(1,0){30}}
\put(0,70){\line(1,0){30}} \put(0,90){\line(1,0){30}}
\end{picture}}
\put(90,0){
\begin{picture}(50,100)(0,0)
\put(0,-1){\line(1,0){30}} \put(0,1){\line(1,0){30}}
\put(0,20){\line(1,0){30}} \put(0,50){\line(1,0){30}}
\put(0,70){\line(1,0){30}} \put(0,90){\line(1,0){30}}
\end{picture}}
\put(70,0){
\begin{picture}(20,100)(0,0)
\put(0,-10){\framebox(20,110){}} \put(0,-10){\line(1,1){20}}
\put(0,0){\line(1,1){20}} \put(0,10){\line(1,1){20}}
\put(0,20){\line(1,1){20}} \put(0,30){\line(1,1){20}}
\put(0,40){\line(1,1){20}} \put(0,50){\line(1,1){20}}
\put(0,60){\line(1,1){20}} \put(0,70){\line(1,1){20}}
\put(0,80){\line(1,1){20}} \put(0,90){\line(1,1){10}}
\put(10,-10){\line(1,1){10}}
\end{picture}}
\end{picture}}
\end{picture}
%
%
%
\begin{picture}(200,120)(0,0)
\put(20,5){
\begin{picture}(120,70)(0,0)
\put(-20,40){$+$}
\put(20,0){
\begin{picture}(30,100)(0,0)
\put(7,7){$\phi\hspace{-.05cm}^-$} \put(-20,7){in}
\put(20,-5){\line(0,1){30}} \put(0,10){\line(4,3){20}}
\put(0,10){\line(4,-3){20}} \put(10,70){$\psi$} \put(-20,70){out}
\put(20,45){\line(0,1){50}} \put(0,70){\line(4,5){20}}
\put(0,70){\line(4,-5){20}}
\end{picture}}
\put(120,0){
\begin{picture}(30,100)(0,0)
\put(1,6){$\phi\hspace{-.05cm}^-$} \put(23,7){out}
\put(0,-5){\line(0,1){30}} \put(20,10){\line(-4,3){20}}
\put(20,10){\line(-4,-3){20}} \put(3,70){$\psi$} \put(23,70){in}
\put(0,45){\line(0,1){50}} \put(20,70){\line(-4,5){20}}
\put(20,70){\line(-4,-5){20}}
\end{picture}}
\put(40,0){
\begin{picture}(30,100)(0,0)
\put(0,-1){\line(1,0){30}} \put(0,1){\line(1,0){30}}
\put(0,20){\line(1,0){30}} \put(0,50){\line(1,0){30}}
\put(0,70){\line(1,0){30}} \put(0,90){\line(1,0){30}}
\end{picture}}
\put(90,0){
\begin{picture}(50,100)(0,0)
\put(0,-1){\line(1,0){30}} \put(0,1){\line(1,0){30}}
\put(0,20){\line(1,0){30}} \put(0,50){\line(1,0){30}}
\put(0,70){\line(1,0){30}} \put(0,90){\line(1,0){30}}
\end{picture}}
\put(70,0){
\begin{picture}(20,100)(0,0)
\put(0,-10){\framebox(20,110){}} \put(0,-10){\line(1,1){20}}
\put(0,0){\line(1,1){20}} \put(0,10){\line(1,1){20}}
\put(0,20){\line(1,1){20}} \put(0,30){\line(1,1){20}}
\put(0,40){\line(1,1){20}} \put(0,50){\line(1,1){20}}
\put(0,60){\line(1,1){20}} \put(0,70){\line(1,1){20}}
\put(0,80){\line(1,1){20}} \put(0,90){\line(1,1){10}}
\put(10,-10){\line(1,1){10}}
\end{picture}}
\end{picture}}
\end{picture}
%
\caption{ Including the negative energy wave functions
$\phi^-_\pi$ in the RGM overlaps.} \label{Negative diagrams}
\end{figure}
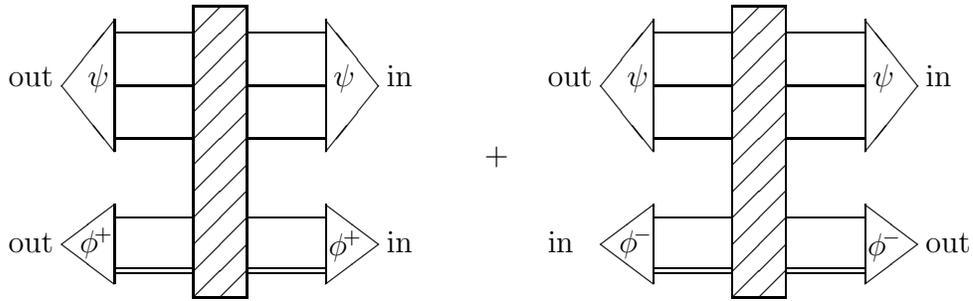

For simplicity the negative energy wave function $\phi^-_\pi$ was
not included in the set of diagrams of Fig. \ref{RGM diagrams}. As
we said, only $\phi^+_\pi$ were included. There an incoming meson
is a Dirac bra that couples to the right side of the RGM diagrams
and an outgoing meson is a Dirac ket that couples to the left side
of the RGM diagrams. To obtain the full set of the RGM overlaps,
containing both $\phi^+_\pi$ and $\phi^-_\pi$ it is sufficient to
notice that negative energy wave functions $\phi^-_\pi$ must
couple as Dirac bras to the right side of the diagram when they
are outgoing and as Dirac kets to the left side when they are
incoming and that they can only come in pairs either of two
$\phi^+_\pi$s or two $\phi^-_\pi$s. The only possible topologies
are depicted in Fig. \ref{Negative diagrams}. Including
$\phi^-_\pi$s amounts to double the number of diagrams of
Fig.\ref{RGM diagrams}. This has been studied in detail in Refs.
\cite{pi-pi,BicRib2,Bicudo4}.

In general $\langle \mathcal{O}^{\mbox{\tiny RGM}}_{\pi-B} \rangle
= \langle \mathcal{O}^{\mbox{\tiny RGM}}_{\pi-B} \rangle_{colour}
\langle \mathcal{O}^{\mbox{\tiny RGM}}_{\pi-B} \rangle_ {spin
\times flavour \times momentum}$. Therefore we need to disentangle
the various contributions for $\mathcal{O}^{\mbox{\tiny
RGM}}_{\pi-B}$ in the spin$\times$fla\-vour$\times$mo\-men\-tum
space. To achieve this it is enough to notice two things: 1 - In
the case of the nucleon, $B=N$, the spin gets mixed with flavour
due to the fact that the spin$\times$flavour nucleon wave function
is given by $(F_{spin}F_{flavour}+D_{spin}D_{flavour})/\sqrt{2}$.
There is no such mixing in the case of the $\Delta$. Therefore in
order to disentangle the spin$\times$flavour is enough to work
separately for the F and D spin components. 2 - The flavour gets
mixed with momentum due to the presence of negative energy
amplitudes. This is a curious effect deserving some explanation.
>From the discussion in the introduction of this section we can
have either two $\phi^+$s or two $\phi^-$s. When going from
$\phi^+$ to $\phi^-$ what was a ket state becomes a bra state and
vice versa. That is, the position of the two $\phi$s in the trace
get interchanged when we go from $\phi^+$ to $\phi^-$. In general
this is enough to make the flavour traces to distinguish between
the two cases. The same effect would have occurred in the spin
case would it not be for the fact the pion has no total spin.
Again, in order to disentangle flavour$\times$momentum it is
enough to give the flavour results discriminated in $\phi^+\phi^+$
and $\phi^-\phi^-$.

Having the above considerations in mind we now proceed to the separate
evaluation of colour, spin, flavour and momentum contributions to
$\mathcal{O}^{\mbox{\tiny RGM}}_{\pi-B}$.

\subsection{Spin and colour contributions to $\mathcal{O}^{\mbox{\tiny RGM}}_{\pi-B}$}

The spin and colour contributions to RGM are independent of the
energy-spin degree of freedom and therefore can be evaluated for
the diagrams of Fig. \ref{RGM diagrams}. The colour contributions
to $\mathcal{O}^{\mbox{\tiny RGM}}_{\pi-B}$ are given in Table
\ref{tab:colour}. As for the spin, we give in Table \ref{tab:spin}
the results for $\mathcal{O}^{\mbox{\tiny RGM}}_{\pi-B}$
discriminated by F and D spin components for the nucleon case.
Notice that this overlaps are given in terms of  the Salpeter
$V_\pi^{++}$, $V_\pi^{+-}$, $V_\rho^{++}$, $V_{N_F}$, $V_{N_D}$
and $V_{\Delta}$ discussed in Section \ref{bound state}. For
instance, the annihilation overlap $\langle A_{15} \rangle$ is
proportional to the $V_\pi^{+-}$ that couples the positive energy
component of the $\pi$ to its negative energy component. The quark
exchange overlaps $\langle V_{14} \rangle$, $\langle V_{25}
\rangle$ and $\langle V_{23} \rangle$, include a linear
combination of the spin singlet $V_\pi^{++}$ and of the spin
triplet $V_\rho^{++}$.

\begin{table}
\caption{Color contributions.}
\label{tab:colour}
\begin{tabular}{ccccc}
\hline
\hline
\makebox[40pt]{$\langle E \rangle \, , \, \langle T_i \rangle$} &
\makebox[20pt]{$\langle V_{14} \rangle$} &
\makebox[20pt]{$\langle V_{25} \rangle$} &
\makebox[20pt]{$\langle V_{23} \rangle$} &
\makebox[20pt]{$\langle A_{15} \rangle$} \\
    \hline
    $\frac{1}{3}$ & $\frac{4}{9}$ & $-\frac{2}{9}$ & $-\frac{2}{9}$ & $\frac{4}{9}$ \\
    \hline
\end{tabular}
\vspace{0.05\textwidth}
\end{table}

\begin{table}
\caption{Spin contributions}
\label{tab:spin}
\begin{tabular}{c l l l}
\hline
\hline
    \multicolumn{1}{c}{} & \multicolumn{2}{c}{$N \pi$} &
\multicolumn{1}{c}{$\Delta \pi$} \\
    \multicolumn{1}{c}{} & \multicolumn{1}{c}{F} &
\multicolumn{1}{c}{D} \\ \hline
    $\langle E \rangle \, , \, \langle T_i \rangle$
& $\frac{1}{2}\delta_{ss'}$
& $\frac{1}{2}\delta_{ss'}$
& $\frac{1}{2}\delta_{ss'}$ \smallskip \\
    $\langle V_{14} \rangle$
& $-\frac{3}{8}V_\pi^{++}(\trD{p},\trD{q})\delta_{ss'}$
& $-\frac{3}{8}V_\pi^{++}(\trD{p},\trD{q})\delta_{ss'}$
& $-\frac{3}{8}V_\pi^{++}(\trD{p},\trD{q})\delta_{ss'}$ \smallskip \\
    $\langle V_{25} \rangle$
& $-\frac{1}{4}V_{N^F}(\trD{p},\trD{q})$
& $-\frac{1}{4}V_{N^D}(\trD{p},\trD{q})$
& $-\frac{1}{4}V_\Delta(\trD{p},\trD{q})$ \smallskip \\
    $\langle V_{23} \rangle$
& $\frac{1}{4}V_{N_F}(\trD{p},\trD{q})$
& $\frac{1}{4}V_{N_D}(\trD{p},\trD{q})$
& $\frac{1}{4}V_\Delta(\trD{p},\trD{q})$ \smallskip \\
    $\langle A_{15} \rangle$
& $-\frac{3}{8}V_\pi^{+-}(\trD{p},\trD{q})\delta_{ss'}$
& $-\frac{3}{8}V_\pi^{+-}(\trD{p},\trD{q})\delta_{ss'}$
& $-\frac{3}{8}V_\pi^{+-}(\trD{p},\trD{q})\delta_{ss'}$  \smallskip \\
\hline
\end{tabular}
\end{table}

\subsection{Flavour contributions}

\begin{table}[h]
\caption{Flavor contributions. $\tau_N$, $\tau_\pi$ are,
respectively, the isospin generators acting in I= 1/2 and 1
isospin wave functions. The I represents the identity operator in
flavor space}
\label{tab:flavourN}
\begin{tabular}{c c l l l l l l l}
\hline
\hline
    \multicolumn{2}{c}{} & \multicolumn{3}{c}{F} &&
\multicolumn{3}{c}{D} \\
    \multicolumn{2}{c}{} & \multicolumn{1}{c}{$\phi^+\phi^+$} &&
\multicolumn{1}{c}{$\phi^-\phi^-$} && \multicolumn{1}{c}{$\phi^+\phi^+$} &&
\multicolumn{1}{c}{$\phi^-\phi^-$} \\ \hline
    $\langle E \rangle \, , \, \langle T_i \rangle$
&\ & $\frac{1}{2}I$ && $\frac{1}{2}I$ &&
$\frac{1}{2}I+\frac{2}{3}\taut{N}$ &&
$\frac{1}{2}I-\frac{2}{3}\taut{N}$ \smallskip \\
    $\langle V_{14} \rangle$
&\ & $\frac{1}{2}I$ && $\frac{1}{2}I$ &&
$\frac{1}{2}I+\frac{2}{3}\taut{N}$ &&
$\frac{1}{2}I-\frac{2}{3}\taut{N}$ \smallskip \\
    $\langle V_{25} \rangle$
&\ & $\frac{1}{2}I$ && $\frac{1}{2}I$ &&
$\frac{1}{2}I+\frac{2}{3}\taut{N}$ &&
$\frac{1}{2}I-\frac{2}{3}\taut{N}$ \smallskip \\
    $\langle V_{23} \rangle$
&\ & $\frac{1}{2}I+\taut{N}$ && $\frac{1}{2}I-\taut{N}$ &&
$\frac{1}{2}I-\frac{1}{3}\taut{N}$ &&
$\frac{1}{2}I+\frac{1}{3}\taut{N}$ \smallskip \\
    $\langle A_{15} \rangle$
&\ & $\frac{1}{2}I$ && $\frac{1}{2}I$ &&
$\frac{1}{2}I-\frac{2}{3}\taut{N}$ &&
$\frac{1}{2}I+\frac{2}{3}\taut{N}$ \\
\hline
\end{tabular}
\end{table}

\begin{table}[h]
\caption{Flavor contributions. $\tau_\pi$ and $\tau_\Delta$ are,
respectively, the isospin generators acting in I= 1 and 3/2
isospin wave functions. The I represents the identity operator in
flavor space}
\label{tab:flavourDelta}
\begin{tabular}{c c l l l}
\hline
\hline
    \multicolumn{2}{c}{} & \multicolumn{3}{c}{$\Delta \pi$} \\
    \multicolumn{2}{c}{} & \multicolumn{1}{c}{$\phi^+\phi^+$} &&
\multicolumn{1}{c}{$\phi^-\phi^-$} \\ \hline
    $\langle E \rangle \, , \, \langle T_i \rangle$
&\ & $\frac{1}{2}I+\frac{1}{3}\taut{\Delta}$ &&
$\frac{1}{2}I-\frac{1}{3}\taut{\Delta}$ \smallskip \\
    $\langle V_{14} \rangle$
&\ & $\frac{1}{2}I+\frac{1}{3}\taut{\Delta}$ &&
$\frac{1}{2}I-\frac{1}{3}\taut{\Delta}$ \smallskip \\
    $\langle V_{25} \rangle$
&\ & $\frac{1}{2}I+\frac{1}{3}\taut{\Delta}$ &&
$\frac{1}{2}I-\frac{1}{3}\taut{\Delta}$ \smallskip \\
    $\langle V_{23} \rangle$
&\ & $\frac{1}{2}I+\frac{1}{3}\taut{\Delta}$ &&
$\frac{1}{2}I-\frac{1}{3}\taut{\Delta}$ \smallskip \\
    $\langle A_{15} \rangle$
&\ & $\frac{1}{2}I-\frac{1}{3}\taut{\Delta}$ &&
$\frac{1}{2}I+\frac{1}{3}\taut{\Delta}$ \\
\hline
\end{tabular}
\end{table}

In the case of the flavour diagrams with different energy-spin
asymptotic states for the mesons contribute differently. As we
have said the existence of negative energy-spin $\phi^-$ amounts
to the doubling of diagrams of Fig. \ref{RGM diagrams} each of
them giving rise to two diagrams, one with $\phi^+\phi^+$ mesonic
asymptotic states and another with $\phi^-\phi^-$ mesonic
asymptotic states. No diagrams with $\phi^+\phi^-$ are allowed.

We give the isospin traces in Tables \ref{tab:flavourN} and
\ref{tab:flavourDelta} for $N\pi$ and $\Delta \pi$ in terms of the
$\phi^\pm$ content of the diagrams $\langle E \rangle$, $\langle
T_i \rangle$, $\langle V_{14} \rangle$, $\langle V_{23} \rangle$
and $\langle A_{15} \rangle$. As in the spin case we must
distinguish between the $F$ and $D$ flavour components because the
nucleon wave function mixes this components with the correspondent
$F$ and $D$ spin. So the net result of considering negative energy
amplitudes for the mesons is to double the entries in Table
\ref{tab:flavourN} for each of the flavours $F$ and $D$. In this
respect the $\Delta \pi$ case is simpler and we can give final
flavour results for $\phi^+\phi^+$ and $\phi^-\phi^-$.

Next we can assemble the results of Tables \ref{tab:colour},
\ref{tab:spin}, \ref{tab:flavourN} and \ref{tab:flavourDelta} to
obtain the final $\pi-B$ results for
colour$\times$spin$\times$flavour$\times$momentum presented in
Tables \ref{tab:Nucleon} and \ref{tab:Delta}.

%
\begin{table}
\caption{Nucleon Total Contribution. The upper (lower) sum/subtraction signs
are associated with $\phi^+\otimes\phi^+$ ($\phi^-\otimes\phi^-$).
$E^{\pm}=E_B\pm E_\pi$, with $E_B$ and $E_\pi$ being the energy of the incoming
baryon and pion}
\label{tab:Nucleon}
\begin{tabular}{c l l}
\hline
\hline
    $\langle E \rangle \, , \, \langle T_i \rangle$
&& $\frac{1}{4}(I\pm\frac{2}{3}\taut{N})(2T_1+2T_2+T_5-E^{\pm})$ \smallskip \\
    $\langle V_{14} \rangle$
&& $-\frac{1}{4}(I\pm\frac{2}{3}\taut{N})\Vpi{++}{p}{q}$ \smallskip \\
    $\langle V_{25} \rangle$
&& $-\frac{1}{8}I\ \Vpi{++}{p}{q}-\frac{1}{8}(I\pm\frac{4}{3}
\taut{N})\Vro{++}{p}{q}$ \smallskip \\
    $\langle V_{23} \rangle$
&& $\frac{1}{16}(I\pm2\taut{N})\Vpi{++}{p}{q}+\frac{1}{16}
(I\mp\frac{2}{3}\taut{N})\Vro{++}{p}{q}$ \smallskip \\
    $\langle A_{15} \rangle$
&& $\frac{1}{4}(I\mp\frac{2}{3}\taut{N})\Vpi{+-}{p}{q}$ \\
\hline
\end{tabular}
\end{table}
%
%
\begin{table}[tbp]
\caption{Delta Total Contribution. The upper (lower) sum/subtraction signs
are associated with $\phi^+\otimes\phi^+$ ($\phi^-\otimes\phi^-$).}
\label{tab:Delta}
\begin{tabular}{c l l}
\hline
\hline
$\langle E \rangle \, , \, \langle T_i \rangle$
&& $\frac{1}{4}(I\pm\frac{2}{3}\taut{\Delta})(2T_1+2T_2+T_5-E^\pm)$ \smallskip \\
$\langle V_{14} \rangle$
&& $-\frac{1}{4}(I\pm\frac{2}{3}\taut{\Delta})\Vpi{++}{p}{q}$ \smallskip \\
$\langle V_{25} \rangle$
&& $-\frac{1}{4}(I\pm\frac{2}{3}\taut{\Delta})\Vro{++}{p}{q}$ \smallskip \\
$\langle V_{23} \rangle$
&& $\frac{1}{8}(I\pm\frac{2}{3}\taut{\Delta})\Vro{++}{p}{q}$ \smallskip \\
$\langle A_{15} \rangle$
&& $\frac{1}{4}(I\mp\frac{2}{3}\taut{\Delta})\Vpi{+-}{p}{q}$ \\
\hline
\end{tabular}
\end{table}

Finally integrating in the internal momenta $p$, $q$ and $k$ we
get for $\mathcal{O}^{\mbox{\tiny RGM}}_{\pi-N}$
 {\samepage
\begin{equation} \label{eq:Ncont}
\begin{split}
\mathcal{O}^{\mbox{\tiny RGM}}_{\pi-N}&=
\int\frac{d^3p}{(2\pi)^3}\frac{d^3q}{(2\pi)^3}\frac{d^3k}{(2\pi)^3} \\
&\left\{ \Psi(p,k)\phi^+(p) \frac{1}{4}A_1(2T_1+2T_2+T_5-E^+)
\Psi(p,k) \phi^+(p)\, \delta^3(\trD{p}-\trD{q})+ \right. \\
&+ \Psi(p,k)\Psi(p,k)\frac{1}{4} \left(A_1 V_\pi^{++} + A_{-1}
V_\pi^{+-}\right)
\phi^+(q) \phi^+(q) + \\
& + \Psi(p,k)\phi^+(p)
\frac{1}{8}\left(I\ V_\pi^{++}+ A_2 V_\rho^{++}\right)
\Psi(q,k)\phi^+(q) + \\
& + \Psi(p,k)\phi^+(k)
\frac{1}{16}\left(A_3 V_\pi^{++}+ A_{-1} V_\rho^{++}\right)
\Psi(q,k) \phi^+(k) + \\
& + \Psi(p,k)\phi^-(p) \frac{1}{4}A_{-1}(2T_1+2T_2+T_5-E^-)
\Psi(p,k) \phi^-(p)\, \delta^3(\trD{p}-\trD{q})+ \\
& + \Psi(p,k)\Psi(p,k) \frac{1}{4}\left(A_{-1} V_\pi^{++} + A_1
V_\pi^{+-}\right)
\phi^-(q) \phi^-(q) + \\
& + \Psi(p,k)\phi^-(p)
\frac{1}{8}\left(I\ V_\pi^{++}+ A_{-2} V_\rho^{++} \right)
\Psi(q,k)\phi^-(q) + \\
&\left. + \Psi(p,k)\phi^-(k) \frac{1}{16}\left(A_{-3} V_\pi^{++}+
A_1 V_\rho^{++}\right) \Psi(q,k) \phi^-(k) \right\} \ ,
\end{split}
\end{equation}
}
with $A_n=I+\frac{2}{3}n\taut{N}$.

Similarly for the $\pi \Delta - \pi \Delta$ overlaps we get
{\samepage
\begin{equation} \label{eq:Dcont}
\begin{split}
\mathcal{O}^{\mbox{\tiny RGM}}_{\pi-\Delta}&=
\int\frac{d^3p}{(2\pi)^3}\frac{d^3q}{(2\pi)^3}\frac{d^3k}{(2\pi)^3} \\
&\left\{ \Psi_\Delta(p,k)\phi^+(p) \frac{1}{4}A_1
(2T_1+2T_2+T_5-E^+)
\Psi_\Delta(p,k) \phi^+(p)\, \delta^3(\trD{p}-\trD{q})+ \right. \\
&+ \Psi_\Delta(p,k)\Psi_\Delta(p,k)
\left(\frac{1}{4}A_1 V_\pi^{++} + \frac{1}{4}A_{-1} V_\pi^{+-}\right)
\phi^+(q) \phi^+(q) + \\
& + \Psi_\Delta(p,k)\phi^+(p)
\left(\frac{1}{4} A_1 V_\rho^{++} \right)
\Psi_\Delta(q,k)\phi^+(q) + \\
& + \Psi_\Delta(p,k)\phi^+(k) \left(\frac{1}{8}A_1
V_\rho^{++}\right)
\Psi_\Delta(q,k) \phi^+(k) + \\
& + \Psi_\Delta(p,k)\phi^-(p) \frac{1}{4}A_{-1}
(2T_1+2T_2+T_5-E^-)
\Psi_\Delta(p,k) \phi^-(p)\, \delta^3(\trD{p}-\trD{q})+ \\
&+ \Psi_\Delta(p,k)\Psi_\Delta(p,k) \left(\frac{1}{4}A_{-1}
V_\pi^{++} + \frac{1}{4}A_1 V_\pi^{+-}\right)
\phi^-(q) \phi^-(q) + \\
& + \Psi_\Delta(p,k)\phi^-(p)
\left( \frac{1}{4} A_{-1} V_\rho^{++} \right)
\Psi_\Delta(q,k)\phi^-(q) + \\
&\left. + \Psi_\Delta(p,k)\phi^-(k)
\left( \frac{1}{8}A_{-1} V_\rho^{++}\right)
\Psi_\Delta(q,k) \phi^-(k) \right\} \ .
\end{split}
\end{equation}}

It is a matter of a simple calculation, to rewrite Eqs.
(\ref{eq:Ncont}) and (\ref{eq:Dcont}) in order to bring forward
the explicit dependence on the matrix elements $\mathcal{H}_\pi$,
$\mathcal{H}_N$ and $\mathcal{H}_\Delta$.

As an example let us consider the $\pi-N$ case. We use the
following general identity
\begin{equation}
\begin{split}
\int\frac{d^3p}{(2\pi)^3}&\int\frac{d^3p'}{(2\pi)^3}\ \
\chi(\trD{p})(2T+V_\pi^{++}(\trD{p},\trD{p}')+V_\pi^{+-}(\trD{p},\trD{p}'))
\zeta(\trD{p}') = \\
&= \frac{1}{2} \int\frac{d^3p}{(2\pi)^3}\int\frac{d^3p'}{(2\pi)^3}
\left[
\begin{array}{cc}
\chi(\trD{p}), &  \chi(\trD{p})
\end{array}
\right]
\mathcal{H}_\pi
\raisebox{-6pt}{$
 \left[
 \begin{array}{c}
 \zeta(\trD{p}') \\
 \zeta(\trD{p}')
 \end{array}\right]$} \ ,
\end{split}
\end{equation}
with $\chi$ and $\xi$ being arbitrary functions, together with the
definition of $\mathcal{H}_N$ and $\mathcal{H}_\Delta$ (see Eqs.
(\ref{compact bound state})), to cast Eq. (\ref{eq:Ncont}) as
{\samepage
\begin{equation} \label{xxx}
\begin{split}
\mathcal{O}^{\mbox{\tiny RGM}}_{\pi-N}&=
\sum_{\alpha=\pm}\int\frac{d^3p}{(2\pi)^3}\frac{d^3q}{(2\pi)^3}\frac{d^3k}{(2\pi)^3}
\\
&\left\{ -\Psi(p,k)\phi^\alpha(p) \frac{E^\alpha}{4}
\left(I+\alpha\frac{2}{3}\taut{N}\right)\Psi(p,k)
\phi^\alpha(p)\, \delta^3(\trD{p}-\trD{q})+ \right. \\
& + \frac{1}{8}\left[
\begin{array}{cc}
\Psi(p,k)\Psi(p,k), & \Psi(p,k)\Psi(p,k)
\end{array}
\right]
\mathcal{H}_\pi \left(I - \alpha \frac{2}{3}\taut{N}\right)
\left[
\begin{array}{c}
\phi^\alpha(q) \phi^\alpha(q) \\
\phi^\alpha(q) \phi^\alpha(q)
\end{array}\right] + \\
&+ \Psi(p,k)\Psi(p,k)\frac{1}{3}
\left(2\mathcal{H}_N-\mathcal{H}_\Delta \right)
\alpha\frac{2}{3}\taut{N}\, \phi^\alpha(q) \phi^\alpha(q) + \\
& + \Psi(p,k)\phi^\alpha(p)
\frac{1}{6}\left( \mathcal{H}_N I +
\mathcal{H}_\Delta \alpha \frac{2}{3}\taut{N} \right)
\Psi(q,k)\phi^\alpha(q) + \\
& \left. + \Psi(p,k)\phi^\alpha(k) \frac{1}{12}\left(\mathcal{H}_N
I + (3\mathcal{H}_N-2\mathcal{H}_\Delta)\alpha\frac{2}{3}\taut{N}
\right) \Psi(q,k) \phi^\alpha(k) \right\} \ .
\end{split}
\end{equation}
}

Finally we use the spectral decomposition of $\mathcal{H}_\pi$,
$\mathcal{H}_N$ and $\mathcal{H}_\Delta$,
\begin{equation}
\begin{split}
H_{\pi }=&\sum_{\pi^*} \sigma _{3}\ \left[
\begin{array}{c}
| \phi ^{+}_{\pi^*} \rangle \\
| \phi ^{-}_{\pi^*} \rangle
\end{array}
\right] \ M_{\pi^*}\ \left[ \langle \phi ^{+}_{\pi^*} |\ \langle
\phi ^{-}_{\pi^*} |\right]
\ \sigma _{3}+ \\
& \hspace{3cm} +
\sigma _{3}\ \left[
\begin{array}{c}
| \phi ^{-}_{\pi^*} \rangle \\
| \phi ^{+}_{\pi^*} \rangle
\end{array}
\right] \ M_{\pi^*}\ \left[ \langle \phi ^{-}_{\pi^*} |\ \langle
\phi ^{+}_{\pi^*} |\right]
\ \sigma _{3}\, , \\
H_{N}=& \sum_{N^*} | \psi_{N^*} \rangle
M_{N^*} \langle \psi_{N^*}|\, , \\
H_{\Delta}=& \sum_{\Delta^*}| \psi_{\Delta^*} \rangle M_{\Delta^*}
\langle \psi_{\Delta^*}| \, ,
\end{split}
\label{closure}
\end{equation}
which are a direct consequence of Eq. (\ref{compact bound state}),
to obtain for $\mathcal{O}^{\mbox{\tiny RGM}}_{\pi-B}$
\begin{equation} \label{eq:IN}
\begin{split}
\mathcal{O}^{\mbox{\tiny RGM}}_{\pi-N}&= \sum_{\alpha=\pm} \left\{
-I_0^\alpha \frac{E^\alpha}{4}
\left(I+\alpha\frac{2}{3}\taut{N}\right) + \frac{1}{8} I_4 M_\pi
\left(I - \alpha \frac{2}{3}\taut{N}\right)
I_5^\alpha+ \right. \\
& + \frac{1}{3} I_1 \left(2M_N-M_\Delta \right)
\alpha\frac{2}{3}\taut{N} I_2^\alpha + \frac{1}{6} I_3^\alpha
\left( M_N I + M_\Delta \alpha \frac{2}{3}\taut{N} \right)
I_3^\alpha + \\
& \left. + \frac{1}{12} I_6 \left(M_N I +
(3M_N-2M_\Delta)\alpha\frac{2}{3}\taut{N} \right) I_6 \right\} \ ,
\end{split}
\end{equation}
with
\begin{equation}
\label{eq:integrals}
\begin{split}
I_0^\pm &= \int \frac{d^3p}{(2\pi)^3}\ \Psi_B(p) \Psi_B(p) \phi^\pm(p) \phi^\pm(p), \\
I_1 &= \int \frac{d^3p}{(2\pi)^3}\ \Psi_B(p) \Psi_B(p)
\Psi_{B'}(p), \
I_2^\pm = \int \frac{d^3p}{(2\pi)^3}\ \phi^\pm(p) \phi^\pm(p) \Psi_{B'}(p), \\
I_3^\pm &= \int \frac{d^3p}{(2\pi)^3}\ \Psi_B(p) \phi^\pm(p)
\Psi_{B'} , \
I_4 = \int \frac{d^3p}{(2\pi)^3}\ \Psi_B(p) \Psi_B(p) (\phi^+(p) -\phi^-(p)), \\
I_5^\pm &= \int \frac{d^3p}{(2\pi)^3}\ \phi^\pm(p) \phi^\pm(p)
(\phi^+(p) -\phi^-(p)) , \ I_6 =  \sqrt{\frac{3}{f_\pi^2 m_\pi}}
\int \frac{d^3p}{(2\pi)^3} \Psi_B(p) \Psi_{B'}(p) .
\end{split}
\end{equation}
We can deal with $\pi-\Delta$ along similar lines. Here we only
quote the result:
\begin{equation} \label{eq:Idelta}
\begin{split}
\mathcal{O}^{\mbox{\tiny RGM}}_{\pi-\Delta}&= \sum_{\alpha=\pm}
\left\{ -I_0^\alpha \frac{E^\alpha}{4}
\left(I+\alpha\frac{2}{3}\taut{\Delta}\right) + \frac{1}{8} I_4
M_\pi \left(I - \alpha \frac{2}{3}\taut{\Delta}\right)
I_5^\alpha+ \right. \\
& + \frac{1}{3} I_1 \left(2M_N-M_\Delta \right)
\alpha\frac{2}{3}\taut{\Delta} I_2^\alpha + \frac{1}{6} I_3^\alpha
M_\Delta \left( I + \alpha
\frac{2}{3}\taut{\Delta} \right) I_3^\alpha + \\
& \left. + \frac{1}{12} I_6 M_\Delta \left(I +
\alpha\frac{2}{3}\taut{\Delta} \right) I_6 \right\} \ .
\end{split}
\end{equation}

We have succeeded in writing $\mathcal{O}^{\mbox{\tiny
RGM}}_{\pi-B}$ in terms of hadron masses and geometric overlaps
only. Any explicit dependence on the microscopic quark kernel $K$
as disappeared. The above formulas (\ref{eq:IN}) and
(\ref{eq:Idelta}) are kernel independent. In the next section we
are going to evaluate the geometrical overlaps $I$'s of Eq.
(\ref{eq:integrals}) in the point-like limit approximation.

\begin{figure}
\centering
\includegraphics[width=0.50\textwidth]{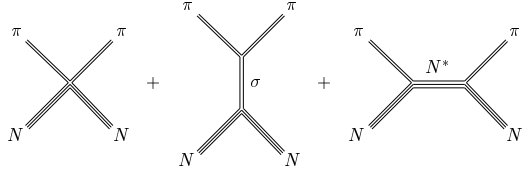}
\caption{ Classes of diagrams in $\pi-B$ scattering}
\label{XYH}
\end{figure}

\begin{figure}
\begin{center}
\begin{picture}(160,60)(0,0)
\put(50,-15){
\begin{picture}(120,70)(0,0)
\put(-50,20){$Z^q=$}
\put(80,0){
\begin{picture}(30,100)(0,0)
\put(-60,20){\line(-1,0){40}}
\put(20,20){\line(-1,0){40}}
\put(0,39){\line(-1,0){20}}
\put(0,41){\line(-1,0){20.5}}
\put(0,60){\line(-1,0){20}}
\end{picture}}
\put(30,30){
\begin{picture}(20,100)(0,0)
\put(0,0){\line(-1,-1){10}}
\put(0,0){\line(1,1){30}}
\put(20,0){\line(1,-1){10}}
\put(20,-1){\line(1,1){10}}
\put(19.5,1){\line(1,1){10}}
\multiput(-1.5,0)(5,0){4}{\line(1,0){2}}
\put(0,0){\circle*{3}}
\put(20,0){\circle*{3}}
\end{picture}}
\put(80,40){
\begin{picture}(30,100)(0,0)
\put(2,5){$\phi\hspace{-.05cm}^+$}
\put(0,-5){\line(0,1){30}}
\put(20,10){\line(-4,3){20}}
\put(20,10){\line(-4,-3){20}}
\end{picture}}
\put(10,0){
\begin{picture}(20,100)(0,0)
\put(0,26){$\Gamma^{q q}$}
\put(48,26){$\Gamma^{\bar q q}$}
\put(33,33){$V$}
\end{picture}}
\end{picture}}
\end{picture}
%
\begin{picture}(150,60)(0,0)
\put(35,-15){
\begin{picture}(120,70)(0,0)
\put(-35,20){$+$}
\put(20,0){
\begin{picture}(30,100)(0,0)
\put(0,20){\line(-1,0){40}}
\put(80,20){\line(-1,0){40}}
\put(0,39){\line(-1,0){20}}
\put(0,41){\line(-1,0){20.5}}
\put(0,60){\line(-1,0){20}}
\end{picture}}
\put(-20,40){
\begin{picture}(30,100)(0,0)
\put(7,8){$\phi\hspace{-.05cm}^-$}
\put(20,-5){\line(0,1){30}}
\put(0,10){\line(4,3){20}}
\put(0,10){\line(4,-3){20}}
\end{picture}}
\put(30,30){
\begin{picture}(20,100)(0,0)
\put(0,0){\line(-1,-1){10}}
\put(0,-1){\line(-1,1){10}}
\put(0,1){\line(-1,1){10}}
\put(20,0){\line(1,-1){10}}
\put(20,0){\line(-1,1){30}}
\multiput(-1.5,0)(5,0){4}{\line(1,0){2}}
\put(0,0){\circle*{3}}
\put(20,0){\circle*{3}}
\end{picture}}
\put(10,0){
\begin{picture}(20,100)(0,0)
\put(-5,26){$\Gamma^{q \bar q}$}
\put(48,26){$\Gamma^{q q}$}
\put(33,33){$V$}
\end{picture}}
\end{picture}}
\end{picture}
\end{center}
\caption{ $Z^q$ annihilation of a meson in a quark line}
\label{Z diagrams}
\end{figure}
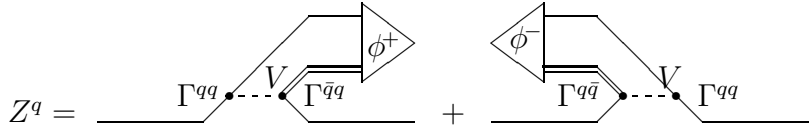

\section{The point like limit}
\label{just contact}

In this paper we follow closely the work done in the $\pi-\pi$
case \cite{pi-pi}. There it was shown, for any microscopic model
complying with chiral symmetry, that in order to calculate
$\pi-\pi$ scattering lengths it is sufficient to consider the
point-like limit,
\begin{equation}
\label{point-like} S(k)\longrightarrow 1\ .
\end{equation}
In this limit only contact diagrams contribute to $\pi-\pi$ $T$
matrix. This is clearly seen if we compute
\cite{BicRib2,Bicudo,Bicudo4}
the coupling of a meson to a single
quark line with the $Z^q$ annihilation diagrams depicted in Fig.
\ref{Z diagrams}. In this limit, the vertices present in Fig.
\ref{Z diagrams}, which represent the integral
\begin{equation}
\begin{split}
Z^q_{s_1,s_2}(p) =& \sum_{s_3,s_4} \int\frac{d^3 p'}{(2\pi)^3}
\Gamma^{qq}_{s_1,s_3}(p,p')
V(p-p') \Gamma^{\bar q q}_{s_2,s_4}(p',p) \phi^+_{s_3,s_4}(p') + \\
& + \sum_{s_3,s_4} \int\frac{d^3 p'}{(2\pi)^3} \Gamma^{q \bar
q}_{s_1,s_3}(p,p') V(p-p') \Gamma^{q q}_{s_2,s_4}(p',p)
\phi^-_{s_3,s_4}(p') \ ,
\end{split}
\end{equation}
vanish. Assumption (\ref{point-like}) implies the vanishing of the
products of vertices,
\begin{equation}
\begin{split}
\Gamma^{q q} \Gamma^{\bar q q} \rightarrow 0  \ , & \hspace{.5cm}
\Gamma^{q \bar q} \Gamma^{q q} \rightarrow 0 \ , \\
\Gamma^{\bar q \bar q} \Gamma^{\bar q q} \rightarrow 0  \ , &
\hspace{.5cm} \Gamma^{q \bar q} \Gamma^{\bar q \bar q} \rightarrow
0 \ .
\end{split}
\label{vanish}
\end{equation}
In the class of models embodied by Eq. (\ref{heff}) $Z^q$ (and
$Z^{\bar q}$) are the microscopic sources of intermediate hadronic
exchange so that, in this limit, diagrams which do not belong to
the class of contact diagrams vanish. When leaving this limit both
classes of diagrams contribute to $T_{\pi-\pi}$ matrix, but then
chiral symmetry forces cancellations among diagrams belonging to
each of these two different classes, so that the $T_{\pi-\pi}$
transition matrix remains formally invariant. This cancellation is
clearly seen in the sigma model calculation of $\pi-\pi$
scattering \cite{Alfaro} where the X shaped diagram of the contact
four $\pi$ coupling and the H shaped diagram of the sigma meson
exchange miraculously cancel and this provides the small physical
result. This result is encoded by the existence of the Adler
zeroes \cite{Adler1} which are present whenever a low energy pion
couples to hadrons. In references \cite{pi-pi,Bicudo1} we have
shown this cancellation for microscopic models defined for the
class of Hamiltonians of Eq. (\ref{heff}). As a byproduct it was
also shown that the transition matrix $T_{\pi-\pi}$ is formally
independent of the quark microscopic kernel $K$. In fact
$T_{\pi-\pi}$ was shown to depend explicitly only on physical
masses $m_\pi$ and wave-function normalizations $f_\pi$. So that
any dependence on the quark kernel $K$ is buried into the masses
and normalization constants dependence on $K$. Therefore we were
allowed, in order to find the $T_{\pi-\pi}$ formula in terms of
$m_\pi$ and $f_\pi$ to choose a kernel consistent to
$S(k)\rightarrow 1$. In this paper we follow the prescription of
$S(k)\rightarrow 1$ and therefore we will only consider contact
diagrams.

The integrals $I$ of Eq. (\ref{eq:integrals}) are particularly
simple to do in the point like limit. In Eqs. (\ref{eq:IN}) and
(\ref{eq:Idelta}), the only relevant combinations of $I$'s we need
to be concerned with, have, up to the first order in $\xi=(\phi^+
-\phi^-)/2a$, the following point like limits
\begin{equation}
\begin{split} \label{pointlike}
&I_0^+ +I_0^- = \int\frac{d^3p}{(2\pi)^3}\ \Psi_B\Psi_B\phi^+\phi^+ +
\int \frac{d^3p}{(2\pi)^3}\ \Psi_B\Psi_B\phi^-\phi^- \\
& = \frac{3}{f_\pi^2 M_\pi}\int \frac{d^3p}{(2\pi)^3}\ \Psi_B\Psi_B S^2 +
\frac{4}{3} f_\pi^2 M_\pi\int \frac{d^3p}{(2\pi)^3}\ \Psi_B\Psi_B\xi^2 \\
& \stackrel{S\rightarrow 1}{\longrightarrow} \frac{3}{f_\pi^2
M_\pi}\int\frac{d^3p}{(2\pi)^3}\ \Psi_B\Psi_B
= \frac{3}{f_\pi^2 M_\pi}\\
&I_0^+ -I_0^- = \int \frac{d^3p}{(2\pi)^3}\ \Psi_B\Psi_B\phi^+\phi^+ -
\int \frac{d^3p}{(2\pi)^3}\ \Psi_B\Psi_B\phi^-\phi^- \\
& = 4 \int \frac{d^3p}{(2\pi)^3}\ \Psi_B\Psi _B S\, \frac{1}{2a}(\phi^+ - \phi^-) \\
& \stackrel{S\rightarrow 1}{\longrightarrow}
\frac{1}{\mathcal{N}^2}4\int\frac{d^3p}{(2\pi)^3}\ S\,
\frac{1}{2a}(\phi^+ - \phi^-) = \frac{1}{\mathcal{N}^2} \\ \\
&\{ I_1 I_2^+ + I_1 I_2^-, I_1 I_2^+ - I_1 I_2^-\}
\stackrel{S\rightarrow 1}{\longrightarrow} \{\frac{3}{f_\pi^2 M_\pi},\frac{1}{\mathcal{N}^2}\} \\
&\{ I_3^+ I_3^+ + I_3^- I_3^-, I_3^+ I_3^+ - I_3^- I_3^-\}
\stackrel{S\rightarrow 1}{\longrightarrow} \{\frac{3}{f_\pi^2 M_\pi},\frac{1}{\mathcal{N}^2}\} \\
&\{ I_6 I_6 + I_6 I_6, I_6 I_6 - I_6 I_6\} \stackrel{S\rightarrow
1}{\longrightarrow} \{\frac{3}{f_\pi^2 M_\pi},0\} \\
&\{ I_4 I_5^+ + I_4 I_5^-, I_4 I_5^+ - I_4 I_5^-\}
\stackrel{S\rightarrow 1}{\longrightarrow}
\{\mathcal{O}(M_\pi),\mathcal{O}(M_\pi)\}
\end{split}
\end{equation}

In order to derive the limits of Eq. (\ref{pointlike}), Eqs.
(\ref{wavefunction}), (\ref{eq:norm}) and (\ref{Bapprox}) were
used. The last entry of Eq. (\ref{pointlike}) presents
combinations of the $I$'s which are of the order $M_\pi$. Because
we are interested in scattering lengths evaluated to order zero in
$M_\pi$ these combinations will not be considered.

Using Eq. (\ref{pointlike}) we can further simplify
$\mathcal{O}^{\mbox{\tiny RGM}}_{\pi-B}$ noticing that due to energy
conservation $E^\pm_{\pi B}= M_B \pm M_\pi$, the following
combination
\begin{equation}
3\;I\;\frac{- E^+_{\pi B} -E^-_{\pi B} + 2M_B}{8 f_\pi^2 M_\pi}=\mathcal{O}(M_\pi) \ .
\label{isoscalar cancellation}
\end{equation}
is of order $M_\pi$. Therefore it will be discarded for the same
reasons as before.

To zero order in $M_\pi$, only the $\taut{B}$ terms survive,
and the total overlap kernel
$\mathcal{O}^{\mbox{\tiny RGM}}_{\pi-B}$  turns out to be quite simple,
\begin{equation}
\mathcal{O}^{\mbox{\tiny RGM}}_{\pi-B} = \frac{4M_N-2M_\Delta}{9
\mathcal{N}^2}\taut{B}+\mathcal{O}(M_\pi ) \ .
\end{equation}

We finally get the desired scattering lengths, see Eq. (\ref{eq:aBorn}),
in a compact notation,
\begin{equation}  \label{tautau}
a_{\pi B}= -\frac{M_\pi}{2 \pi}\frac{4M_N-2M_\Delta}{9
\mathcal{N}^2} \tau_{B}\cdot\tau_\pi+\mathcal{O}(M_\pi^2 )\ .
\end{equation}
The scattering lengths vanish in the chiral limit of a vanishing
quark mass $M_\pi\rightarrow 0$, and this complies with the Adler
zero \cite{Alfaro,Adler1} .

\section{Numerical results and Conclusion}
\label{conclusion}

We have a free parameter, the $N$ and $\Delta$ normalization that
$\mathcal{N}$ needs to be fixed. To this effect we can use the
experimental value for the $a_{\pi N}\{I=1/2\}$ scattering length,
\begin{equation}
a_{\pi N}^{exp}\{I=1/2\} = (0.171 \pm 0.005) M_\pi^{-1} \ ,
\end{equation}
to fix $\mathcal{N}$. We obtain $\mathcal{N}=1580 MeV^{-3/2}$,
corresponding to a nucleon size of $0.4$ fm. Equipped with this
value we predict the following values for

%
%

\begin{equation}
\begin{split}
a_{\pi N}\{I=3/2\}&= -0.086 M_\pi^{-1} \quad  [a_{\pi
N}^{exp}\{I=3/2\}= -(0.088\pm 0.004) M_\pi^{-1}
] \ , \\
a_{\pi \Delta}\{I=1/2\}&= 0.429 M_\pi^{-1} \ , \\
a_{\pi \Delta}\{I=3/2\}&= 0.172 M_\pi^{-1} \ , \\
a_{\pi \Delta}\{I=5/2\}&= -0.258  M_\pi^{-1} \ .
\end{split}
\end{equation}
Although it is difficult o measure directly the $\pi - \Delta$
scattering lengths because the $\Delta$ is quite unstable, the
$\pi - \Delta$ coupling may be determined indirectly in the same
way as the $N - \Delta$ scattering lengths are determined in Ref.
\cite{Garcilazo}. We stress that the QM provides the correct
framework to derive the $\pi - \Delta$ scattering lengths. For
instance it would be hard to derive the $\pi - \Delta$ scattering
lengths in the framework of effective hadronic models.

In particular we predict that the different scattering lengths of
the $\pi$ in s-wave baryons are exactly proportional with
remarkable integer factors,
\begin{equation}
a_{\pi N,I=\frac{1}{2}} : a_{\pi N,I=\frac{3}{2}} : a_{\pi
\Delta,I=\frac{1}{2}}  : a_{\pi \Delta,I=\frac{3}{2}}  : a_{\pi
\Delta,I=\frac{5}{2}} = 2 : -1 : 5 : 2 : -3 \ .
\end{equation}

\par
The work of G. M. Marques is supported by Funda\c c\~ao para a
Ci\^encia e a Tecnologia under the grant SFRH/BD/984/2000.

\appendix

\section{Traces} \label{traces}
In this appendix the computation of the color, spin and flavor
traces are detailed. There are different equivalent methods to
compute these standard traces.

\subsection{color}
The color traces are the most straightforward to calculate. The
baryon color wave-function is a simple $\epsilon^{abc}/ \sqrt{6}$
and the meson a $\delta^{ab} / \sqrt{3}$. The interaction has the
well known structure $\lambda^{ab}/2$. For instance the pion
Bethe-Salpeter color factor 4/3 of Eq. (\ref{bspi}) is calculated
in the following way
\begin{equation}
\frac{\delta_{ij}}{\sqrt{3}}\frac{\lambda^a_{ik}}{2}
\frac{\delta_{kl}}{\sqrt{3}}\frac{\lambda^b_{jl}}{2}=
\frac{1}{12}Tr\{\trD{\lambda}\cdot\trD{\lambda}\}= \frac{4}{3} \,
.
\end{equation}
For the baryons the color factor -2/3 of Eqs. (\ref{Vsimpl1}),
(\ref{Vsimpl2}) and (\ref{Vsimpl3}) comes from
\begin{equation}
\frac{\epsilon_{ijk}}{\sqrt{6}}\frac{\lambda^a_{il}}{2}
\frac{\epsilon_{lmk}}{\sqrt{6}}\frac{\lambda^b_{jm}}{2}=
\frac{1}{24}(Tr\{\trD{\lambda}\}\cdot Tr\{\trD{\lambda}\}-
Tr\{\trD{\lambda}\cdot\trD{\lambda}\})= -\frac{2}{3}
\end{equation}

In Table \ref{tab:colour} we present the results for the diagrams
of Fig. \ref{RGM diagrams}.

\subsection{spin}

For the spin our method relies on the Pauli matrices $\sigma_i$.

The spin structures of $\pi$, $N$ and $\Delta$ wave functions,
together with the spin vector structure $\rho$, later needed in
the calculations, can be written as
\begin{eqnarray} \label{eq:spinstruc}
& \Sigma^\pi =\left(i \sigma_2/\sqrt{2}\right)_{ab};
\hspace{1cm}
\Sigma^\rho_{\mathbf{s}} = \left[\trD{\sigma}i \sigma_2/\sqrt{2}\right]_{ab}
\cdot\trD{v}_{\mathbf{s}};
\hspace{1cm}
\Sigma^\Delta_{\mathbf{s}} = \left[\trD{\sigma}i \sigma_2/\sqrt{2}\right]_{ab}
\cdot \trD{w}_{c\mathbf{s}}; & \nonumber \\
& \Sigma^{N^F}_{\mathbf{s}} = \left(i\sigma_2/\sqrt{2}\right)_{ab}\delta_{c\mathbf{s}};
\hspace{1cm}
\Sigma^{N^D}_{\mathbf{s}} = (1/\sqrt{3})\left[\trD{\sigma}i\sigma_2/\sqrt{2}\right]_{ab}
\cdot\trD{\sigma}_{c\mathbf{s}} & \nonumber \\
\end{eqnarray}
where $a$, $b$ and $c$ stand for the quarks individual spin,
whereas $\mathbf{s}$ stands for the \textbf{total} spin of the
meson or baryon under consideration. For example, we have
\begin{equation}
\left[
\begin{array}{cc}
\uparrow & \downarrow
\end{array}
\right] \Sigma^\pi
\raisebox{-6pt}{$
\left[
\begin{array}{c}
 \uparrow \\
 \downarrow \\
\end{array}
\right]$} =\frac{1}{\sqrt 2}(\uparrow\downarrow-\downarrow\uparrow).
\end{equation}
The vectors $\trD{v}_\mathbf{s}$ and $\trD{w}_{c\mathbf{s}}$, are given by,
\begin{eqnarray}
& \trD{v}_1=\left(-1/\sqrt{2},-i/\sqrt{2},0\right),
\hspace{10pt}
\trD{v}_0=\left(0,0,-1\right),
\hspace{10pt}
\trD{v}_{-1}=\left(1/\sqrt{2},-i/\sqrt{2},0\right), & \nonumber\\
&\trD{w}_{c\frac{3}{2}}=\trD{v}_1\ \delta_{c\frac{1}{2}},
\hspace{10pt}
\trD{w}_{c\frac{1}{2}}=1/\sqrt{3}\ \trD{v}_1\ \sigma^-_{c\frac{1}{2}}+
\sqrt{2/3}\ \trD{v}_0\ \delta_{c\frac{1}{2}}, & \\
& \trD{w}_{c-\frac{1}{2}}=1/\sqrt{3}\ \trD{v}_{-1}\ \sigma^+_{c-\frac{1}{2}}+
\sqrt{2/3} \ \trD{v}_0\ \delta_{c-\frac{1}{2}},
\hspace{10pt} \trD{w}_{c-\frac{3}{2}}=\trD{v}_{-1}\ \delta_{c-\frac{1}{2}}. &
\nonumber
\end{eqnarray}
They have the following properties,
\begin{equation}
\begin{split}
{v^i_s}^\dag v^j_{s'}&=\delta^{ij}\delta_{s s'}-(J^j J^i)_{s s'}, \\
{w^i_{sc}}^\dag w^j_{cs'}&=\frac{3}{4}\delta^{ij}\delta_{ss'}-
\frac{1}{3}(J^i J^j)_{s s'}+\frac{i}{2}\epsilon^{ijk}J^k_{s s'}.
\end{split}
\end{equation}
where the $J$'s are the spin 1 SU(2) generating matrices in the
case of the $v$'s relation and the spin 3/2 SU(2) generating
matrices in the case of the $w$'s relation. When we contract $i$
with $j$ we have simply
\begin{equation}
{v^i_s}^\dag v^i_{s'} = {w^i_{sc}}^\dag w^i_{cs'} = \delta_{ss'}.
\end{equation}
The importance of the above spin functions lies in the fact that
they map the quark spin content of a given bound state to the
total spin of that bound state. Once this map is achieved it is a
matter of straightforward calculations to obtain the diagrammatic
traces. The results are presented in table \ref{tab:spin
appendix}. For instance for the spin traces which contribute to
the RGM $\langle E \rangle$ and $\langle V_{24} \rangle$ overlap
diagrams the computation is
\begin{equation}
\begin{split}
\langle E \rangle_{spin} &=
Tr\{\Sigma^B_s\cdot{\Sigma^B_{s'}}^\dag\cdot\Sigma^\pi\cdot{\Sigma^\pi}^\dag\}= \\
&=-\frac{1}{2}\delta_{ss'} \ ,
\end{split}
\end{equation}
\begin{equation}
\begin{split}
\langle V_{24} \rangle_{spin} &=
Tr\{\Sigma^B_s\cdot{\Sigma^B_{s'}}^\dag\cdot\Gamma^{qq}
\cdot\Sigma^\pi\cdot{\Sigma^\pi}^\dag\cdot\Gamma^{qq}\}V(\trD{p}-\trD{p}')= \\
&= \frac{1}{2}Tr\{\Sigma^B_s\cdot{\Sigma^B_{s'}}^\dag\cdot\Gamma^{qq}
\cdot\Gamma^{qq}\}V(\trD{p}-\trD{p}')= \\
&= -\frac{3}{8}V_\pi^{++}(\trD{p},\trD{p}')\delta_{ss'} \ .
\end{split}
\end{equation}
where the super-index $B$ stands for the corresponding baryon which may be
the $F$ component of the nucleon, the $D$ component of the nucleon
or the Delta.

\begin{table}
\caption{Spin contributions} \label{tab:spin appendix}
\begin{tabular}{c l}
\hline
\hline
    $\langle E \rangle \, , \, \langle T_i \rangle$
& $Tr\{\Sigma^B_s\cdot{\Sigma^B_{s'}}^\dag\cdot\Sigma^\pi\cdot{\Sigma^\pi}^\dag\}$ \\
    $\langle V_{14} \rangle$
& $Tr\{\Sigma^B_s\cdot{\Sigma^B_{s'}}^\dag\cdot\Gamma^{qq}
\cdot\Sigma^\pi\cdot{\Sigma^\pi}^\dag\cdot\Gamma^{qq}\}V(\trD{p}-\trD{q})$ \\
    $\langle V_{25} \rangle$
& $Tr\{\Sigma^B_s\cdot\Gamma^{qq}\cdot{\Sigma^B_{s'}}^\dag
\cdot\Sigma^\pi\cdot\Gamma^{\bar{q}\bar{q}}\cdot{\Sigma^\pi}^\dag\}V(\trD{p}-\trD{q})$ \\
    $\langle V_{23} \rangle$
& $\frac{1}{2}Tr\{\Sigma^B_s\cdot\Gamma^{qq}\cdot{\Sigma^B_{s'}}^\dag\cdot\Gamma^{qq}\}
V(\trD{p}-\trD{q})$ \\
    $\langle A_{15} \rangle$
& $Tr\{\Sigma^B_s\cdot{\Sigma^B_{s'}}^\dag\cdot\Gamma^{q\bar{q}}
\cdot{\Sigma^\pi}^\dag\cdot\Sigma^\pi\cdot\Gamma^{\bar{q}q}\}V(\trD{p}-\trD{q})$ \\
\hline
\end{tabular}
\end{table}

\subsection{flavor}
In the isospin case we use $\tau_i$ as the Pauli matrices.

For $\pi-N$ we have the cases of isospin 3/2 and 1/2.
\begin{eqnarray}
&&(I=3/2):\;p \pi^+ ,\;\frac{1}{\sqrt{3}}(n \pi^+ + \sqrt{2} p \pi^0)\ ,\;
\frac{1}{\sqrt{3}}(\sqrt{2} n \pi^0 + p \pi^-) ,\; n \pi^-  \nonumber\\
&&
(I=1/2):\; \frac{1}{\sqrt{3}}(\sqrt{2} n \pi^+ - p \pi^0),\;
\frac{1}{\sqrt{3}}(n \pi^0 - \sqrt{2} p \pi^-)
\end{eqnarray}
The isospin contribution of any diagram must be a linear combination of the
identity and $\taut{N}$, where
\begin{equation}
\taut{N}=\frac{1}{2}((\tau_\pi+\tau_N)^2-\tau_\pi^2-\tau_N^2)
=\frac{1}{2}(I(I+1)-2-\frac{3}{4}).
\end{equation}
The computation of these contributions was done for each diagram
and each multiplet element. The results are presented in Table
\ref{tab:flavour appendix}.

In the same way for $\Delta-\pi$ we have isospin 5/2, 3/2 and 1/2.
\begin{eqnarray}
&&(I=5/2):\nonumber\\
&&\Delta^{++} \pi^+,\;
\frac{1}{\sqrt{5}}(\sqrt{2} \Delta^{++} \pi^0 + \sqrt{3} \Delta^+ \pi^+)
,\;\frac{1}{\sqrt{10}}(\Delta^{++} \pi^- + \sqrt{6}
\Delta^+ \pi^0 + \sqrt{3} \Delta^0 \pi^+) \nonumber\\
&&\frac{1}{\sqrt{10}}(\sqrt{3} \Delta^{+} \pi^- +
\sqrt{6} \Delta^0 \pi^0 + \Delta^- \pi^+),\;
\frac{1}{\sqrt{5}}(\sqrt{3} \Delta^0 \pi^- +
\sqrt{2} \Delta^- \pi^0),\;
\Delta^{-} \pi^- \nonumber\\
&&(I=3/2):\nonumber\\
&&\frac{1}{\sqrt{5}}(\sqrt{3} \Delta^{++} \pi^0 - \sqrt{2} \Delta^+ \pi^+),\;
\frac{1}{\sqrt{15}}(\sqrt{2} \Delta^{++} \pi^- + \Delta^+ \pi^0 -
\sqrt{8} \Delta^0 \pi^+),\nonumber\\
&&\frac{1}{\sqrt{15}}(\sqrt{8} \Delta^{+} \pi^- - \Delta^0 \pi^0 -
\sqrt{2} \Delta^- \pi^+),\;
\frac{1}{\sqrt{5}}(\sqrt{2} \Delta^0 \pi^- - \sqrt{3} \Delta^- \pi^0)\nonumber\\
&&(I=1/2):\nonumber\\
&&\frac{1}{\sqrt{6}}(\sqrt{3} \Delta^{++} \pi^- - \sqrt{2} \Delta^+ \pi^0 +
\Delta^0 \pi^+),\;
\frac{1}{\sqrt{6}}(\Delta^{+} \pi^- - \sqrt{2} \Delta^0 \pi^0 + \sqrt{3}
\Delta^- \pi^+) \nonumber\\
\end{eqnarray}
In this case the isospin contributions must be a linear combination of the
identity, $\taut{\Delta}$ and $(\taut{\Delta})^2$, once more where
\begin{equation}
\taut{\Delta}=\frac{1}{2}((\tau_\pi+\tau_\Delta)^2-\tau_\pi^2-\tau_\Delta^2)
=\frac{1}{2}(I(I+1)-2-\frac{15}{4}).
\end{equation}
Table \ref{tab:flavour appendix} includes also the isospin
structures of each diagram for $\Delta-\pi$ scattering.

\begin{table}[h]
\caption{Flavor contributions. $\tau_N$, $\tau_\pi$ and
$\tau_\Delta$ are, respectively, the isospin generators acting in
I= 1/2, 1 and 3/2 isospin wave functions. The I represents the
identity operator in flavor space} \label{tab:flavour appendix}
\begin{tabular}{c c l l l l l}
\hline
\hline
    \multicolumn{2}{c}{} & \multicolumn{3}{c}{$N \pi$} &&
\multicolumn{1}{c}{$\Delta \pi$} \\
    \multicolumn{2}{c}{} & \multicolumn{1}{c}{F} &&
\multicolumn{1}{c}{D} && \\ \hline
    $\langle E \rangle \, , \, \langle T_i \rangle$
&\ & $\frac{1}{2}I$ && $\frac{1}{2}I+
\frac{2}{3}\taut{N}$ & &$\frac{1}{2}I+
\frac{1}{3}\taut{\Delta}$ \smallskip \\
    $\langle V_{14} \rangle$
&\ & $\frac{1}{2}I$ && $\frac{1}{2}I+
\frac{2}{3}\taut{N}$ && $\frac{1}{2}I+
\frac{1}{3}\taut{\Delta}$ \smallskip \\
    $\langle V_{25} \rangle$
&\ & $\frac{1}{2}I$ && $\frac{1}{2}I+
\frac{2}{3}\taut{N}$ && $\frac{1}{2}I+
\frac{1}{3}\taut{\Delta}$ \smallskip \\
    $\langle V_{23} \rangle$
&\ & $\frac{1}{2}I+\taut{N}$ && $
\frac{1}{2}I-\frac{1}{3}\taut{N}$ && $
\frac{1}{2}I+\frac{1}{3}\taut{\Delta}$ \smallskip \\
    $\langle A_{15} \rangle$
&\ & $\frac{1}{2}I$ &&
$\frac{1}{2}I-\frac{2}{3}\taut{N}$ && $\frac{1}{2}I-
\frac{1}{3}\taut{\Delta}$ \\
\hline
\end{tabular}
\end{table}


\begin{thebibliography}{00}
%
\bibitem{Weinberg}
S.~Weinberg,
Phys.\ Rev.\ Lett.\  {\bf 17} (1966) 616.
%
\bibitem{pi-pi}
P.~Bicudo, S.~Cotanch, F.~Llanes-Estrada, P.~Maris, E.~Ribeiro and A.~Szczepaniak,
Phys.\ Rev.\ D {\bf 65} (2002) 076008
[arXiv:hep-ph/0112015].
%
\bibitem{Bicudo1}
P.~Bicudo,
Phys.\ Rev.\ C {\bf 67} (2003) 035201.
%
\bibitem{Ribeiro2} J. Ribeiro, Nuc. Phys. A {\bf 689},247c (2001).
%
\bibitem{MIT}
A. Chodos, R.L. Jaffe, K. Johnson, C.B. Thorn, and V.F. Weisskopf,
Phys. Rev. D {\bf 9}, 3471 (1974); A. Chodos, R.L.
Jaffe, K. Johnson, and C.B. Thorn, {\it ibid} {\bf 10}, 2599 (1974);
T.A. de Grand, R.L. Jaffe, K. Johnson, and J. Kiskis, {\it ibid}
{\bf 12}, 2060 (1975).
%
\bibitem{CT}
A. Chodos and C.B. Thorn, Phys. Rev. D. {\bf 12}, 2733
(1975); T. Inoue and T. Maskawa, Prog. Theor. Phys. {\bf 54}, 1833
(1975).
%
\bibitem{Vento}
V. Vento, M. Rho and G.E. Brown, Nucl. Phys. {\bf  A345}, 413 (1980).
%
\bibitem{CBMorig}
S. Th\'eberge, A.W. Thomas, and G.A. Miller, Phys.
Rev. D {\bf 22}, 2838 (1980); {\bf 23}, 2106(E) (1981); A.W. Thomas,
S. Th\'eberge, and G.A. Miller, {\it ibib} {\bf 24}, 216 (1981).
%
\bibitem{Nambu}
Y. Nambu and J. Jona-Lasinio,
Phys.\ Rev.\  {\bf 124}, 246 (1961);
Phys.\ Rev.\  {\bf 122}, 345 (1961).
%
\bibitem{LeYaou-pot}
A. Le Yaouanc, L. Oliver, O. P\'ene, and J.-C. Raynal,
Phys. Rev. D {\bf 29}, 1233 (1984).
%
\bibitem{LeYaou-mes}
A. Le Yaouanc, L. Oliver, O. P\'ene, and J.-C. Raynal,
Phys. Rev. D {\bf 31}, 137 (1985).
%
\bibitem{Adler2}
S. Adler and A. Davis,
Nucl.\ Phys.\ B {\bf 244}, 469 (1984).
%
\bibitem{BicRib1}
P.~J.~Bicudo and J.~E.~Ribeiro,
Phys.\ Rev.\ D {\bf 42} (1990) 1611.
%
\bibitem{BicRib2}
P.~J.~Bicudo and J.~E.~Ribeiro,
Phys.\ Rev.\ D {\bf 42} (1990) 1625;
{\bf 42} (1990) 1635.
%
\bibitem{Felipe}
F. Llanes-Estrada and S. Cotanch, Phys. Rev. Lett. {\bf 84}, 1102 (2000).
%
\bibitem{Dosch}
H.G. Dosch  Phys. Lett. B {\bf 190}, 177 (1987);
H.G. Dosch and Yu A. Simonov Phys. Lett. B  {\bf 205}, 339 (1988).
%
\bibitem{Bicudo}
P.~J.~Bicudo,
Phys.\ Rev.\ C {\bf 60} (1999) 035209.
%
\bibitem{Bicudo4}
P.~Bicudo and J.~Ribeiro,
Phys.\ Rev.\ C {\bf 55} (1997) 834
[arXiv:nucl-th/9703027].
%
\bibitem{Bicudo8}
P. Bicudo, G. Krein and J. Ribeiro, Phys. Rev. C {\bf 64}, 025202 (2001).
%
\bibitem{Bicudo9}
P. Bicudo, J. Ribeiro and J. Rodrigues Phys. Rev. C {\bf 52}, 2144 (1995).
%
\bibitem{Bicudo10}
P. Bicudo, L.Ferreira, C. Placido and J. Ribeiro, Phys. Rev. C {\bf 56}, 670 (1997).
%
\bibitem{Ribeiro1}
J.~E.~Ribeiro,
Z.\ Phys.\ C {\bf 5} (1980) 27.
%
\bibitem{Lemaire}
S. Lemaire, J. Labarsouque and B. Silvestre-Brac, Nucl. Phys. {\bf A714},
265, (2003).
%
\bibitem{Ceuleneer}
R. Ceuleneer, C. Semay and B. Silvestre-Brac,
J. Phys. G {\bf 22}, 1395 (1996).
%
\bibitem{Bicudo6}
P. Bicudo, J. E. Ribeiro and A. Nefediev, Phys.\ Rev.\ D{\bf 65} :085026 (2002).
%
\bibitem{Ribeiro3} J. E. Ribeiro, Phys. Rev. D {\bf 25}, 2406 (1982);
E. van Beveren, Zeit. Phys. C {\bf 17}, 135 (1982).
%
\bibitem{Deus}J. Dias de Deus and J. Ribeiro Phys. Rev. D {\bf 21}, 1251 (1980).
%
\bibitem{Bicudo7}
P. Bicudo, G. Krein, E. Ribeiro and  J. Villate,  Phys. Rev. D {\bf 45}, 1673 (1992).
%
\bibitem{Alfaro}
V. De Alfaro, S. Fubini, G. Furlan, C. Rosseti,
``Currents in Hadron Physics'' ,
Amsterdam, North-Holland, (1973).
%
\bibitem{Adler1}
S.~L.~Adler,
Phys.\ Rev.\  {\bf 137} (1965) B1022.
%
\bibitem{Garcilazo}
H. Garcilazo, P. Sauer, T. Mizutani and M. Pena,
Phys.\ Rev.\ C {\bf 42} 2315 (1990).
%
\end{thebibliography}
\end{document}